\documentclass{report}

\title{CNN Architecture Comparison for Malware Classification with Limited Training Data}
\author{Noor Abo Hasan}
\usepackage[disable]{todonotes}
\usepackage{xcolor}
\usepackage[disable]{todonotes}
\usepackage{comment}
\usepackage[T1]{fontenc}
\usepackage{booktabs}
\usepackage{helvet}

\usepackage{hyperref}
\usepackage{xurl} 
\usepackage{breakurl}

\usepackage{lipsum}

\usepackage{lineno}

\usepackage{soul}

\usepackage{graphicx}
\usepackage{subcaption}
\usepackage{float} 
\usepackage{tikz}
\usetikzlibrary{arrows.meta, positioning, fit, shapes.geometric}
\usetikzlibrary{positioning, shapes}

\usepackage{natbib}

\usepackage[linesnumbered,ruled,noline,noend]{algorithm2e}
\SetKwComment{Comment}{/* }{ */}
\SetKwInput{KwData}{Input}
\SetKwInput{KwResult}{Output}

\usepackage{amsmath}
\usepackage{amsthm}
\usepackage{amssymb}

\usepackage{listings}
\definecolor{codegreen}{rgb}{0,0.6,0}
\definecolor{codegray}{rgb}{0.5,0.5,0.5}
\definecolor{codepurple}{rgb}{0.58,0,0.82}
\definecolor{grayBackground}{rgb}{0.9,0.9,0.9}
\definecolor{greenBackground}{rgb}{0,0.6,0}
\lstdefinestyle{grayBackground}{
    backgroundcolor=\color{grayBackground},   
    commentstyle=\color{codegreen},
    keywordstyle=\color{magenta},
    numberstyle=\tiny\color{codegray},
    stringstyle=\color{codepurple},
    basicstyle=\ttfamily\footnotesize,
    breakatwhitespace=false,         
    captionpos=b,                    
    keepspaces=true,                 
    numbersep=5pt,                  
    showspaces=false,                
    showstringspaces=false,
    showtabs=false,                  
    tabsize=2
}
\usepackage{listings}
\definecolor{codegreen}{rgb}{0,0.6,0}
\definecolor{codegray}{rgb}{0.5,0.5,0.5}
\definecolor{codepurple}{rgb}{0.58,0,0.82}
\definecolor{grayBackground}{rgb}{0.9,0.9,0.9}
\definecolor{greenBackground}{rgb}{0.8,1,0.8}
\lstdefinestyle{greenBackground}{
    backgroundcolor=\color{greenBackground},   
    commentstyle=\color{codegreen},
    keywordstyle=\color{magenta},
    numberstyle=\tiny\color{codegray},
    stringstyle=\color{codepurple},
    basicstyle=\ttfamily\footnotesize,
    breakatwhitespace=false,         
    captionpos=b,                    
    keepspaces=true,                 
    numbers=left,                    
    numbersep=5pt,                  
    showspaces=false,                
    showstringspaces=false,
    showtabs=false,                  
    tabsize=2
}

\definecolor{lightaqua}{rgb}{0.8, 1.0, 1.0}

\newtheorem*{problemformulation}{Problem Formulation}
\newtheorem*{researchquestion}{Research Question}

\begin{document}
\thispagestyle{empty}

\begin{center}
\begin{Huge} \sffamily
CNN Architecture Comparison for Malware Classification with Limited Training Data
\end{Huge}

\vspace{1em}

\begin{Large} \sffamily
A Comparative Study of Single-Path and Dual-Path CNNs Across Varying Training Data Sizes
\vspace{1em}

Noor Abo Hasan
\end{Large}
\end{center}

\vspace{250pt plus 1pt minus 1pt}
\begin{Large} \sffamily
$ $\bigskip\newline\indent
Department of Computer \newline\indent
and Systems Sciences
\end{Large}
\begin{small} \sffamily
\newline\indent
\begin{minipage}{0.65\textwidth}
Degree project 15 credits\\\indent
Computer and Systems Sciences\\\indent
Degree project at the bachelor level\\\indent
Spring term 2025\\\indent
Supervisor: Mateus de Oliveira Oliveira\\\indent
Swedish title: Bildbaserad Malware-klassificering: En Jämförande Studie av Single-Path och Dual-Path CNN med Varierande Träningsdata
\end{minipage}
\end{small}
\begin{minipage}{0.32\textwidth}
\begin{center}
    \includegraphics[width=0.8
    \textwidth]{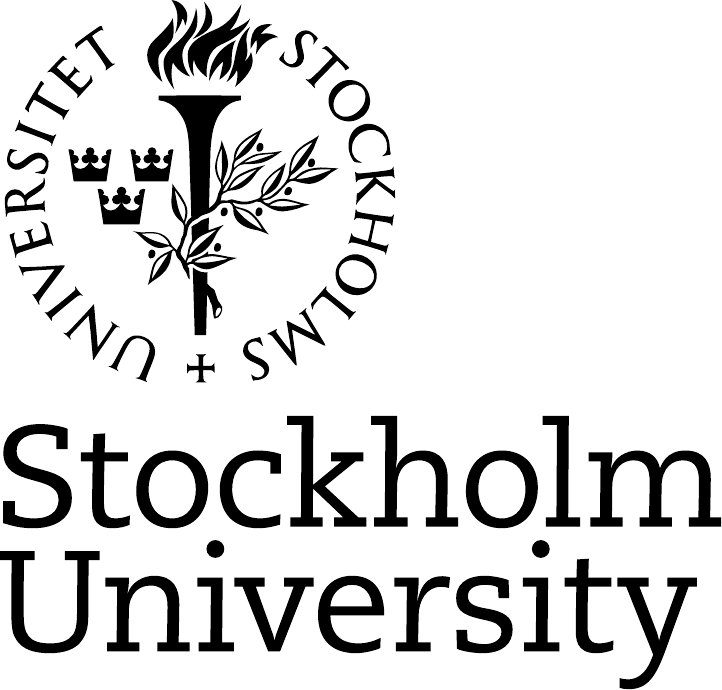}
\end{center}
\end{minipage}
\newpage
\thispagestyle{empty}
\mbox{}
\clearpage

\pagestyle{empty}
\setcounter{page}{1}
\pagenumbering{roman}
\pagestyle{plain}
\thispagestyle{empty}
\noindent
\textbf{DVK Uppsats Program} \\

\noindent 
This thesis was written within the Spring 2025 edition of the 
DVK Uppsats Program. 
\\ 

\noindent
\textbf{Coordination}\\
\\ 
Mateus de Oliveira Oliveira
\\ 

\noindent
{\bf Organization}\\
\\
\noindent
Henrik Bergström\\ 
Peter Idestam-Almquist\\
Mateus de Oliveira Oliveira\\ 
Beatrice Åkerblom

\clearpage
\thispagestyle{empty}
\mbox{}

\chapter*{Abstract}
\thispagestyle{empty}
Traditional malware classification methods struggle to keep pace with rapidly evolving threats. Image-based malware classification offers a promising alternative by converting malware binaries into grayscale images for analysis with deep learning. While effective, organizations often face practical constraints in acquiring sufficient training data, especially for newer threats. This thesis investigates dual-path CNN architectures with intermediate feature fusion as a potential enhancement for malware classification, with particular focus on how varying training data size affects their comparative performance against traditional single-path approaches.

This research implements and compares two models: a baseline single-path using the pre-trained VGG16 architecture and a dual-path model that combines VGG16 with a custom CNN through intermediate feature fusion. Both architectures are evaluated on a binary classification task distinguishing between two malware families (C2Lop.P trojan and Dialplatform.B worm) from the Malimg dataset. Each model is trained across four different training data proportions (20\%, 40\%, 60\%, and 80\%), with three training runs per size to ensure statistical robustness, while being evaluated on complementary test sets of varying sizes.

Though both architectures achieved perfect classification performance across all training percentages, the study revealed significant differences in secondary metrics including training efficiency, robustness to noise perturbation, and architectural complexity. By systematically comparing these architectures under varying data constraints, this thesis contributes insights into the practical trade-offs between architectural complexity and performance. The findings challenge common assumptions about complex architectures, demonstrating that simpler models can offer superior efficiency and robustness without sacrificing classification performance in image-based malware analysis.
\newline\newline
\emph{Keywords:}  Malware Classification, Image-Based Analysis, Dual-Path CNN, Intermediate Feature Fusion, Training Data Efficiency, Data Size Variation, Model Robustness.

\chapter*{Synopsis}
\thispagestyle{empty}

\subsection*{Background}
Malware classification is a critical cybersecurity challenge requiring continuous innovation to counter evolving threats. Traditional detection methods often fail against sophisticated obfuscation techniques, leading researchers to explore alternative approaches. Image-based malware analysis – converting binary files to grayscale images and applying computer vision techniques – has emerged as a promising direction. While Convolutional Neural Networks have demonstrated effectiveness in this domain, practical implementation often faces data constraints, as organizations may have limited malware samples, particularly for newer threats. Understanding how architectural choices affect performance under varying data availability conditions is therefore crucial for developing practical malware classification systems.

\subsection*{Problem}
Image-based malware classification using CNNs has shown promising results, but standard single-path architectures may not optimally capture diverse visual patterns. Dual-path architectures with intermediate feature fusion offer a potential solution, yet their data efficiency—the ability to maintain performance with limited training data—remains unexplored. Current research lacks systematic comparisons of how these architectures perform across varying training data sizes, limiting practical implementation in data-constrained environments. This thesis addresses this knowledge gap by evaluating which architectural approach maintains better performance when training data is limited.

\subsection*{Research Question}
This thesis addresses the following research question: How does varying training data size affect the comparative performance of single-path and dual-path CNN architectures with intermediate feature fusion in malware classification using grayscale image representations?

By investigating this question, the research aims to determine which architectural approach offers better resilience to data limitations, a common constraint in real-world security applications.

\subsection*{Method}

The research employs an experimental design comparing two architectures: a single-path CNN using the pre-trained VGG16 model (baseline) and a dual-path CNN that fuses features from VGG16 and a custom CNN. Both models are trained on four different proportions of the available data (20\%, 40\%, 60\%, and 80\%), with three distinct training runs for each percentage to ensure statistical robustness. For each experiment, the dataset is split using stratified sampling, with training data comprising the specified percentage and the complementary portion serving as the test set. The dataset includes samples from C2Lop.P (trojan) and Dialplatform.B (worm) families from the Malimg dataset, with preprocessing including resizing to 224×224 pixels and normalization. Model performance is evaluated using primary classification metrics (accuracy, precision, recall, F1-score) as well as secondary metrics including training efficiency, robustness to noise perturbation, and architectural complexity.

\subsection*{Result}

The experimental results revealed that both single-path and dual-path architectures achieved perfect classification performance (100\% accuracy, precision, recall, and F1-score) across all training data percentages. With primary metrics reaching saturation, secondary performance indicators became critical differentiators between architectures. The single-path architecture demonstrated substantial advantages in computational efficiency, requiring 18-51\% less training time than its dual-path counterpart, with the efficiency gap widening as training data increased. Most surprisingly, the simpler architecture exhibited superior robustness to noise perturbation—maintaining perfect accuracy under high noise conditions where the dual-path model's performance degraded significantly. Analysis of model complexity revealed that despite having slightly fewer total parameters, the dual-path architecture contained substantially more trainable parameters and required 35\% more storage space.

\subsection*{Discussion}

The study's findings, particularly the saturation of primary performance metrics alongside clear differences in secondary indicators, guide the main discussion. While both architectures solved the classification task, the analysis focuses on interpreting the trade-offs revealed. The single-path model's pronounced advantages in training efficiency and noise robustness, despite equal accuracy, suggest that for this specific, highly separable malware classification scenario, architectural simplicity provides greater practical value and resilience than the more complex dual-path fusion approach. These insights have implications for selecting resource-efficient and reliable models in real-world security contexts where task difficulty and data quality may vary.

\chapter*{Acknowledgement}
\thispagestyle{empty}
I would like to thank my supervisor, Mateus de Oliveira Oliveira, for his invaluable guidance and mentorship throughout this work. His feedback and support were instrumental in keeping this research on track. I also extend my appreciation to Peter Idestam-Almquist for serving as a reviewer and providing thorough, constructive comments.

\newpage
\setcounter{page}{1}
\pagenumbering{roman}
\thispagestyle{plain}
\tableofcontents

\listoffigures
\addcontentsline{toc}{chapter}{List of Figures}

\listoftables
\addcontentsline{toc}{chapter}{List of Tables}

\chapter*{List of Abbreviations}

CNN - Convolutional Neural Network \\
GPU - Graphics Processing Unit \\
RGB - Red, Green, Blue (color model) \\
SVM - Support Vector Machine \\
VGG16 - A 16-layer CNN architecture by Visual Geometry Group \\
\addcontentsline{toc}{chapter}{List of Abbreviations}
\addtocontents{toc}{\bigskip}

\chapter{Introduction}
\label{chapter:Introduction}

\pagenumbering{arabic} %
\setcounter{page}{1} %
\pagestyle{plain}

In the digital arms race of cybersecurity, malware authors continuously evolve their tactics to evade detection systems. As signature-based detection falters against polymorphic and obfuscated threats, security researchers have turned toward visual analytics—transforming malware binaries into grayscale images that reveal distinctive structural patterns \citep{nataraj2011malware}. This approach leverages the observation that malware variants within a family share visual signatures even when their code has been substantially modified, potentially providing resilience against obfuscation techniques that defeat traditional detection methods.

Convolutional Neural Networks have emerged as powerful tools for analyzing these malware visualizations, capable of automatically extracting the complex patterns that distinguish malicious code structures. However, conventional single-path CNN architectures may not fully capture the multifaceted visual signatures present in malware images. Dual-path architectures with feature fusion offer a potential solution \citep{Zhang_2022}, yet their data efficiency—the ability to maintain performance with limited training data—remains unexplored. While architectural complexity often implies better performance, simpler models might actually demonstrate advantages in data-limited scenarios, challenging common assumptions in deep learning. As organizations often face constraints in acquiring large malware datasets, understanding how architectural choices affect performance across varying training data sizes becomes critical for practical implementation \citep{Almajed2025}.

\section{Background}

Image-based malware analysis transforms binary executables into visual representations by mapping byte values to pixel intensities, typically arranged in a fixed-width grid. This visualization preserves spatial relationships within the code, revealing distinctive patterns: code regions appear as structured sections with limited grayscale variation, while encrypted or compressed content creates areas of high entropy visualized as noise-like textures \citep{nataraj2011malware}. These visual representations make otherwise complex binary structures accessible to image analysis techniques.

Convolutional Neural Networks excel at hierarchical feature learning—early layers identify basic visual elements while deeper layers assemble these into complex pattern representations. This architecture parallels how malware analysts examine binaries: from identifying low-level structures to recognizing higher-level patterns indicative of malicious functionality \citep{lecun2015deep}. Unlike manual analysis, however, CNNs can automatically discover subtle patterns across thousands of samples without explicit programming of feature extraction rules.

Dual-path CNN architectures diverge from traditional single-path designs by processing input through parallel feature extraction pipelines before integration. This approach enables simultaneous application of complementary feature extractors—such as a pre-trained model capturing general image characteristics alongside a specialized network targeting domain-specific patterns. Feature fusion, the process of combining these distinct feature sets, creates unified representations that incorporate diverse information types before classification \citep{Boulahia2021}.

Intermediate feature fusion represents a strategic integration point where features from parallel CNN pathways are combined while still retaining spatial information. Unlike early fusion (input-level combination) or late fusion (decision-level integration), intermediate fusion allows the network to learn cross-pathway feature relationships that preserve both global context and local details. In practice, this typically involves concatenating or adding feature maps from corresponding layers across paths, creating expanded feature representations that subsequent shared layers can analyze holistically.

Data efficiency in deep learning refers to a model's ability to achieve high performance with limited training data. This characteristic is particularly important in cybersecurity domains where gathering large, labeled malware datasets presents significant challenges. More complex architectures often require more training data to properly tune their parameters, potentially creating a tradeoff between architectural sophistication and data requirements \citep{Sun2017RevisitingUE}. Understanding this relationship is crucial for developing practical malware classification systems that can perform effectively despite data constraints.

Classifying malware into families offers significant advantages beyond simple detection. By recognizing specific family traits, security systems can deploy targeted countermeasures, predict behaviors, and identify relationships between evolving threats. Family classification also enables detection of new variants through visual similarity to known families, even when code signatures have been altered. The feature discrimination required for family differentiation is typically more sophisticated than that needed for binary malware detection, often leading to more robust feature extractors that strengthen overall security capabilities.

\section{Problem}

Image-based malware classification has demonstrated impressive results in controlled research environments, with CNNs achieving high accuracy rates on benchmark datasets \citep{kalash2018}. However, single-path CNN architectures—the dominant approach in current research—process features through a sequential pipeline that may not optimally capture the diverse visual characteristics present in malware images. This architectural limitation becomes increasingly significant as malware creators employ sophisticated obfuscation techniques that alter visual signatures while preserving malicious functionality.

Dual-path CNN architectures offer a potential solution by enabling parallel feature extraction through complementary pathways. A critical aspect of these architectures is intermediate feature fusion—the strategic integration of features before classification decisions. While this approach has shown promise in other computer vision domains, its application to malware classification remains limited. The existing research lacks systematic comparisons between single-path CNNs and dual-path networks with feature fusion for malware family discrimination tasks.

Furthermore, real-world deployment of malware classification systems often faces practical constraints in dataset acquisition, particularly for newer threats where limited samples are available. This raises important questions about architectural data efficiency—how different CNN designs perform under varying training data availability. Current research has not adequately examined whether the potential benefits of dual-path architectures persist when training data is limited, or whether simpler single-path approaches might actually provide more robust performance with smaller datasets. The assumption that more complex architectures invariably yield better results deserves critical examination, particularly in domains with practical data constraints.

This knowledge gap impedes progress toward more robust and practical malware classification systems. Without empirical evidence comparing how architectural alternatives perform across different training data sizes, security researchers lack clear guidance on which approach offers the best performance under realistic data constraints.

\begin{problemformulation}
\label{problemformulationOne} 
\mbox{}\\[1ex]
\textit{Image-based malware classification using CNNs has shown promising results, but standard single-path architectures may miss complementary patterns crucial for distinguishing malware families. While dual-path architectures with intermediate feature fusion offer a potential solution, their comparative performance advantages across varying training data sizes remain inadequately explored, limiting their practical implementation in data-constrained environments.}\\[1ex]

\end{problemformulation}
With malware samples increasing by more than 20 percent in a single year from 696 million to 852 million Windows malware samples in 2023 alone \citep{avtest2023}—architectural improvements that enhance classification performance could offer significant practical value for cybersecurity systems.

\section{Research Question}

This thesis investigates how training data size affects the relative performance of different CNN architectures for malware classification. By comparing single-path and dual-path approaches across varying data availability conditions, this research aims to determine whether the potential benefits of feature fusion persist when training data is limited or if simpler architectures provide more robust performance with smaller datasets.

\begin{researchquestion}
\label{researchQuestion}

How does varying training data size affect the comparative performance of single-path and dual-path CNN architectures with intermediate feature fusion in malware classification using grayscale image representations?

\end{researchquestion}

To address this question, this study will implement an experimental comparison between two architectures: a baseline single-path CNN using the pre-trained VGG16 model and a dual-path CNN that combines VGG16 with a custom CNN through intermediate feature fusion. Both models will be evaluated on a binary classification task distinguishing between C2Lop.P (trojan) and Dialplatform.B (worm) families from the Malimg dataset. Each architecture will be trained using four different portions of the available training data (20\%, 40\%, 60\%, and 80\%), with three distinct training sets generated for each percentage to ensure statistical robustness. Performance will be measured through standard Primary Performance Metrics, complemented by Training Efficiency, Model Complexity Analysis, and Robustness to Noise to provide a multi-dimensional comparison of architectural performance under varying data constraints.

This study focuses specifically on binary classification between two malware families as a controlled experiment to isolate architectural effects. While real-world malware detection typically involves multi-class scenarios across numerous families, this focused approach enables clearer assessment of how architectural choices impact performance under varying data constraints. By examining the four specific training data sizes the investigation provides a comprehensive view of performance scaling behaviors. The use of multiple training runs for each data size enhances statistical validity. For each training percentage, multiple distinct train/test splits were generated using stratified sampling to ensure both statistical robustness and valid performance comparisons across different data availability conditions. Results should be interpreted as insights into architectural robustness under data constraints rather than a comprehensive security solution, particularly given that the Malimg dataset represents a historical snapshot that may not fully reflect contemporary malware characteristics. The use of multiple distinct train/test splits for each condition enhances the reliability of the comparative findings.

\section{Related Work}

To properly contextualize this investigation of how CNN architectures perform across varying training data sizes, it is essential to examine how previous research has approached image-based malware classification and architectural data efficiency.

The visualization of malware as grayscale images was pioneered by \citep{nataraj2011malware}, who demonstrated that malware families exhibit distinctive visual signatures that can be leveraged for classification. Building on this approach, \citep{kalash2018} applied deep learning techniques to malware image classification, establishing CNNs as effective tools for this task and achieving near-perfect accuracy on benchmark datasets.

Transfer learning has emerged as a valuable technique in malware classification, with \citep{Vasan2020IMCFNIM} showing that pre-trained networks like VGG16 can be effectively repurposed for malware analysis despite being originally trained on natural images. This approach leverages general feature extraction capabilities developed on large datasets while adapting to the specific patterns present in malware visualizations.

Research on data efficiency in deep learning has shown that architectural choices significantly impact performance with limited data. \cite{DBLP:journals/corr/abs-1902-07208} demonstrated that complex models often underperform simpler alternatives when training data is scarce, challenging the assumption that more sophisticated architectures always yield better results. These findings suggest that architectural complexity should be matched to data availability, a principle with important implications for malware classification, where labeled datasets may be limited.

Particularly relevant to this thesis, \cite{yan2018detecting} explored a two-stream CNN architecture for malware classification that processed both the raw binary and image representations of malware samples. Their approach demonstrated how combining features from complementary information sources could enhance classification performance, although they focused on different input modalities rather than parallel feature extractors within the same modality.

The work most conceptually similar in architecture is the Dual Convolutional Malware Network (DCMN) by \cite{almasri2020dcmn}, which also combined a pre-trained model (ResNet-50) with a custom CNN path using feature concatenation. However, their study focused on maximizing classification accuracy on the full 25-class Malimg dataset (plus a benign category) using a fixed data split (60\% train). In contrast, this thesis specifically investigates the impact of varying training data percentages (20\% to 80\%) on the comparative efficiency and robustness of a VGG16-based dual-path architecture employing intermediate fusion against a single-path baseline, particularly in a binary classification context where primary metrics saturated.

Despite these advances, systematic investigation of how single-path and dual-path architectures compare across different training data sizes remains limited in malware classification research. This thesis contributes to the field by conducting a controlled comparison between architectural approaches under varying data availability conditions, providing empirical evidence on which approach offers better performance and robustness when training data are limited, a common constraint in real-world security applications.
\section{Results and Considerations}

This research revealed several unexpected and significant insights regarding CNN architectural choices for malware image classification. Both single-path and dual-path architectures achieved perfect classification performance (100\% accuracy, precision, recall, and F1-score) across all training data percentages, indicating that this specific binary classification task was highly separable even with limited data. This performance saturation highlighted the importance of secondary metrics in architectural evaluation.

The single-path architecture demonstrated substantial advantages in computational efficiency, requiring 18-51\% less training time than its dual-path counterpart, with the efficiency gap widening as training data increased. Most surprisingly, the simpler architecture exhibited superior robustness to noise perturbation—maintaining perfect accuracy under high noise conditions where the dual-path model's performance degraded significantly. These findings challenge the common assumption that more complex architectures necessarily yield better results in visual malware classification, suggesting that architectural simplicity may offer both efficiency and robustness benefits without sacrificing performance.

Looking forward, this research opens several promising directions for malware classification systems. Future work should extend this comparison to multi-class scenarios with larger families of malware and explore more challenging datasets where primary performance metrics might not reach saturation. Additionally, investigating hybrid architectures that combine the robustness advantages of single-path models with selective domain-specific features could yield even more effective systems. The demonstrated efficiency of the single-path architecture also suggests potential for deployment in resource-constrained environments such as edge devices used in network security monitoring. As malware continues to evolve, these architectural insights will help security researchers develop detection systems that balance performance, efficiency, and robustness—critical factors in the ongoing effort to counter emerging cyber threats.

\chapter{Extended Background}
\label{chapter:Background}

This chapter provides a comprehensive overview of the key techniques relevant to image-based malware detection. The discussion begins with grayscale malware visualization, a technique for transforming malware binaries into images suitable for analysis, as introduced by \cite{nataraj2011malware}. Next, the chapter introduces Convolutional Neural Networks (CNNs), a powerful class of deep learning models used for image classification, which have been widely applied to malware detection \cite{kalash2018}. Following this, the chapter explores dual-path CNN architectures, which enhance feature extraction by processing data through multiple parallel pathways, as exemplified by the Dual Convolutional Malware Network (DCMN) \citep{almasri2020dcmn}. Finally, the chapter discusses feature fusion techniques, focusing on intermediate fusion, which combines features from different pathways to improve detection accuracy and robustness.

\section{Grayscale Malware Visualization}

\subsection{Overview}

Grayscale malware visualization is a technique for transforming malware binaries into grayscale images that can be analyzed using computer vision techniques \citep{nataraj2011malware}. This approach leverages the observation that malware samples belonging to the same family often exhibit similar visual patterns when their binary code is converted into a grayscale image. By representing malware as images, we can harness the power of deep learning models, such as Convolutional Neural Networks (CNNs), to automatically extract features and classify malware families \citep{Vasan2020IMCFNIM}. This technique offers a compelling alternative to traditional signature-based methods, which often struggle to keep pace with the rapid evolution of malware and its obfuscation tactics \citep{nataraj2011malware}. The core idea is that the structure and patterns within the executable code of a malware sample can be visually represented, allowing for the application of image processing and pattern recognition techniques.

\subsection{Binary-to-Image Conversion: A Detailed Process}

The process of converting a malware binary into a grayscale image is a crucial step in enabling the application of CNNs for malware detection. This conversion involves several key steps, each designed to preserve the essential information contained within the binary file while transforming it into a format suitable for image processing. The specific steps and parameters used in this conversion can significantly impact the resulting image and the performance of subsequent analysis techniques \citep{ Vasan2020IMCFNIM}.

\begin{enumerate}
    \item \textit{Reading the Binary File:} The initial step involves reading the malware binary file as a sequence of bytes. This is typically done using standard file I/O operations in a programming language such as Python or C++. The binary file is treated as a stream of bytes, each representing a value between 0 and 255.

    \item \textit{Mapping Bytes to Pixel Values:} Each byte in the binary file is then mapped to a pixel value in the range of 0 to 255, representing the grayscale intensity. This mapping is straightforward, as the byte values directly correspond to the grayscale intensity levels. A byte with a value of 0 is mapped to a black pixel, while a byte with a value of 255 is mapped to a white pixel. Intermediate values are mapped to different shades of gray.

    \item \textit{Reshaping into a 2D Array:} The sequence of pixel values is then reshaped into a two-dimensional array, forming the grayscale image. The width of the image is often fixed, while the height is determined by the size of the binary file. For example, if the binary file has a size of 100,000 bytes and the desired width of the image is 256 pixels, then the height of the image would be approximately 390 pixels $$(100,000 / 256 \approx 390)$$. The reshaping process ensures that the pixel values are arranged in a grid-like structure, which is essential for image processing operations.
    A visual representation of this conversion process is shown in Figure~\ref{figure:Grayscale}.

\begin{figure}[htbp]
  \centering
  \includegraphics[width=1.2\textwidth]{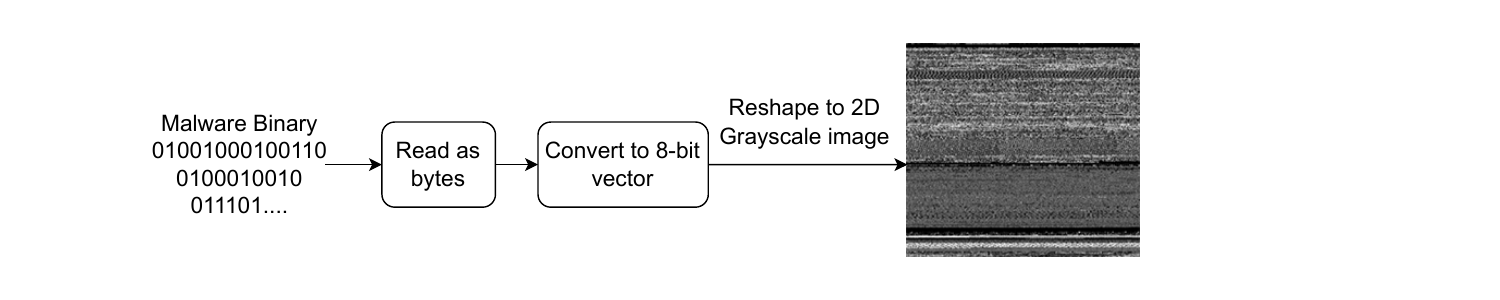}
  \caption{Byte-to-Pixel Mapping for Malware Visualization.}
  \label{figure:Grayscale}
\end{figure}

    \item \textit{Image Normalization (Optional):} In some cases, image normalization techniques may be applied to enhance the contrast and improve the visual quality of the grayscale image. Normalization involves scaling the pixel values to a specific range, such as 0 to 1, which can help to reduce the impact of variations in lighting and improve the performance of subsequent image processing algorithms \citep{zhao2017image}.
\end{enumerate}

The resulting grayscale image provides a visual representation of the underlying binary code, allowing us to identify patterns and textures that may be indicative of malicious behavior. The specific patterns and textures that are visible in the grayscale image depend on the structure and content of the binary file, as well as the specific mapping and reshaping techniques that are used.

\subsection{Factors Influencing Image Representation}

Several factors can influence the visual representation of malware binaries as grayscale images. These factors include the size of the binary file, the choice of image dimensions, and the presence of compression or encryption techniques \citep{Vasan2020IMCFNIM}.

\begin{itemize}
    \item \textit{Binary File Size:} The size of the binary file directly affects the dimensions of the resulting grayscale image. Larger binary files will result in larger images, which may capture more detail but also require more computational resources for processing.

    \item \textit{Image Dimensions:} The choice of image dimensions (width and height) can also impact the visual representation. A wider image may capture more horizontal patterns, while a taller image may capture more vertical patterns. The optimal choice of image dimensions depends on the specific characteristics of the malware dataset and the desired level of detail \citep{zhao2017image}.

    \item \textit{Compression and Encryption:} The presence of compression or encryption techniques in the malware binary can significantly alter the visual representation. Compressed or encrypted binaries may appear as random noise in the grayscale image, making it difficult to identify meaningful patterns \citep{grosse2017}.
\end{itemize}

\subsection{Advantages of Grayscale Visualization}

Grayscale visualization offers several key advantages for malware analysis, making it a valuable tool in the fight against evolving cyber threats:

\begin{itemize}
    \item \textit{Visual Pattern Recognition:} By transforming raw binary code into a visual format, grayscale visualization allows analysts and algorithms to identify recurring patterns and structural similarities within malware families. For example, certain types of packing or obfuscation techniques might create distinctive textures in the resulting images \citep{nataraj2011malware}.

    \item \textit{CNN Compatibility:} Grayscale images are readily compatible with Convolutional Neural Networks (CNNs), which are designed to automatically learn hierarchical features from image data. This eliminates the need for manual feature engineering, allowing the CNN to directly extract relevant characteristics from the visual representation of the malware \citep{Vasan2020IMCFNIM}.

    \item \textit{Simplicity and Efficiency:} Compared to more complex analysis techniques, grayscale visualization is relatively simple to implement and computationally efficient. This makes it a practical choice for analyzing large datasets of malware samples, especially when combined with the parallel processing capabilities of GPUs.

    \item \textit{Feature Extraction Potential:} While CNNs can automatically extract features, grayscale images can also be used as input for traditional image processing techniques, such as edge detection or texture analysis. These techniques can provide complementary information about the malware and potentially improve detection accuracy.

    \item \textit{Value in Limited Data Scenarios:} Grayscale visualization is particularly valuable in contexts where training data is limited—a common constraint in malware analysis due to the challenges of obtaining large, labeled datasets of emerging threats. The visual patterns in malware images often capture family-specific characteristics that remain recognizable even with relatively few samples \citep{Vasan2020IMCFNIM}. This property makes image-based approaches potentially more data-efficient than approaches requiring extensive feature engineering. As demonstrated by \citet{DBLP:journals/corr/abs-1902-07208}, transfer learning with pre-trained vision models can further enhance performance on limited malware datasets by leveraging general feature extraction capabilities developed on larger image datasets. The data efficiency of different CNN architectures when applied to these grayscale representations—particularly how performance scales with training data size—represents an important area of investigation for practical malware classification systems.
    
\end{itemize}

\subsection{Limitations of Grayscale Visualization}

While grayscale visualization offers several advantages, it also has some limitations:

\begin{itemize}
    \item \textit{Loss of Information:} The binary-to-image conversion process inevitably involves some loss of information. The grayscale image represents a simplified view of the underlying binary code, and some of the nuances and complexities of the code may be lost in the conversion \citep{nataraj2011malware}.

    \item \textit{Sensitivity to Obfuscation:} Malware obfuscation techniques can significantly alter the visual representation of the malware binary, making it difficult to identify meaningful patterns. Obfuscation techniques such as code insertion, code reordering, and instruction substitution can change the byte sequence of the binary file, resulting in a different grayscale image \citep{grosse2017}.

    \item \textit{Limited Scalability:} The computational cost of processing large grayscale images can be significant, limiting the scalability of this technique for analyzing large datasets of malware samples.
\end{itemize}

\subsection{Related Techniques}

While grayscale visualization is a popular and effective technique, other approaches to representing malware binaries for analysis exist. One alternative is to use color images, where different bytes or code sections are mapped to different color channels. While this approach may capture more information, it also increases the computational complexity of the model and may not necessarily improve detection accuracy. Another alternative is to represent the malware binary as a sequence of opcodes or API calls, which can then be analyzed using natural language processing techniques \citep{maniriho2023api}. However, this approach requires more sophisticated feature engineering and may not capture the visual patterns that are readily apparent in grayscale images. Grayscale visualization offers simplicity, efficiency, and the ability to capture the essential characteristics of malware binaries in a visual format that can be effectively analyzed by CNNs.

\section{Convolutional Neural Networks (CNNs)}

\subsection{Overview}

Convolutional Neural Networks (CNNs) have revolutionized computer vision, achieving state-of-the-art results in tasks like image classification, object detection, and image segmentation \citep{ goodfellow2016deep}. Their success stems from their ability to automatically learn hierarchical features directly from raw data, a significant advantage over traditional methods that rely on hand-crafted features. CNNs draw inspiration from the hierarchical organization of the visual cortex, mimicking the way neurons respond to specific patterns in their receptive fields \citep{fukushima1980neocognitron}.

While the foundations of CNNs were laid decades ago with models like the Neocognitron \citep{fukushima1980neocognitron} and LeNet-5 \citep{lecun1998gradient}, it was the advent of deeper architectures and larger datasets that truly unlocked their potential. Today, CNNs are ubiquitous in various applications, including image-based malware detection, where they offer a promising approach to classifying malware families based on visual representations of their binary code \citep{kalash2018}.

\subsection{Key Components}
CNNs are built upon several key components that work in concert to extract relevant features and perform image classification. These components include convolutional layers, pooling layers, activation functions, and fully connected layers \citep{lecun2015deep}.

\textbf{CNN Architecture Description:}
A typical CNN architecture starts with an input image, which is then passed through a series of convolutional layers. Each convolutional layer applies a set of learnable filters to the input, extracting features such as edges, textures, and shapes. The output of each convolutional layer is then passed through an activation function, which introduces non-linearity into the network. After several convolutional layers, pooling layers are used to reduce the dimensionality of the feature maps and make the network more robust to variations in the input. Finally, the output of the pooling layers is fed into one or more fully connected layers, which perform the final classification based on the extracted features.

\begin{itemize}
    \item \textit{Convolutional Layers:} Convolutional layers are the fundamental building blocks of CNNs. They consist of a set of learnable filters (also known as kernels) that convolve across the input image, extracting features such as edges, textures, and shapes \citep{ dumoulin2016guide}. Each filter produces a feature map, which represents the response of the filter to different parts of the input image. The filters are learned during the training process through a process called backpropagation, allowing the CNN to automatically adapt to the specific characteristics of the dataset \citep{ goodfellow2016deep}. The convolutional operation can be mathematically expressed as:

    $$
    (f * g)(t) = \int f(\tau)g(t - \tau) d\tau
    $$
    where $f$ is the input image, $g$ is the filter, $*$ denotes the convolution operation, $t$ represents the position in the output feature map, and $\tau$ is the variable of integration representing a position in the input image. This formula essentially computes the weighted sum of the input image values in a neighborhood around position $t$, with the weights given by the filter values. 

    \item \textit{Pooling Layers:} Pooling layers are used to reduce the dimensionality of the feature maps and make the model more robust to variations in the input, such as small shifts or rotations \citep{boureau2010theoretical}. Common pooling operations include max pooling and average pooling, with max pooling being the most widely used. In max pooling, the feature map is divided into non-overlapping regions (typically 2×2 pixels), and the maximum value in each region is selected as the output. This effectively downsamples the feature map while preserving the most important information. By keeping only the most prominent features, pooling helps to reduce computational cost, prevent overfitting, and achieve translation invariance in the network \citep{goodfellow2016deep}.
    
    \item \textit{Activation Functions:} Activation functions introduce non-linearity into the CNN, allowing it to learn complex patterns and relationships in the data \citep{ goodfellow2016deep}. Without non-linear activation functions, the CNN would simply be a linear model, severely limiting its representational capacity. Common choices include ReLU (Rectified Linear Unit), sigmoid, and tanh. ReLU is the most widely used activation function in CNNs due to its simplicity and efficiency, as it helps to alleviate the vanishing gradient problem that can occur with other activation functions \citep{krizhevsky2012imagenet}. The ReLU activation function is defined as:

    $$
    ReLU(x) = max(0, x)
    $$

    where $x$ is the input to the activation function (typically the output of a convolutional or fully connected layer). This function returns $x$ if it is positive, and 0 if it is negative, effectively allowing only positive values to pass through while setting negative values to zero.

    \item \textit{Fully Connected Layers:} Fully connected layers are used to classify the input based on the extracted features. These layers are typically placed at the end of the CNN and consist of a set of neurons that are fully connected to the previous layer. The output of the fully connected layers is a probability distribution over the different classes, representing the model's confidence in each class. The fully connected layers perform a linear transformation followed by a non-linear activation function, similar to a traditional multi-layer perceptron \citep{Rosenblatt1958ThePA, goodfellow2016deep}.
\end{itemize}

\subsection{CNN Architecture Visualization}
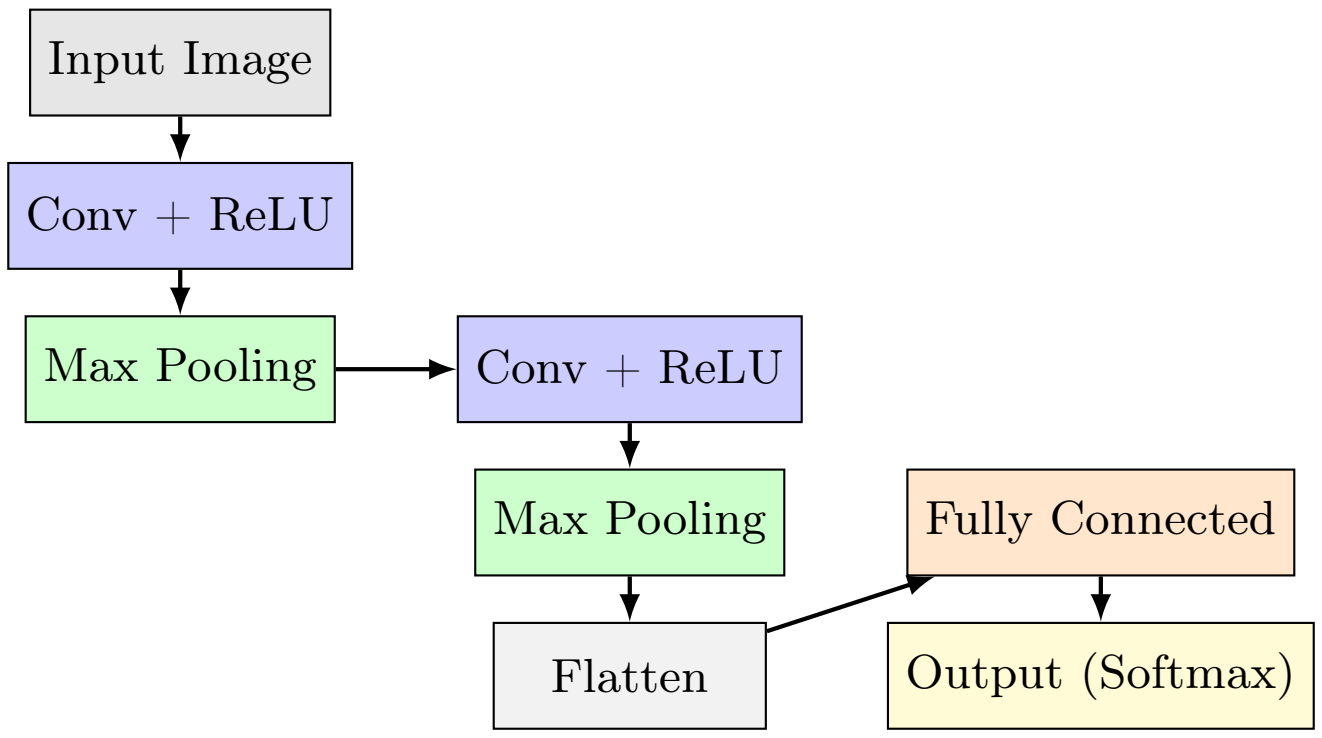
\begin{figure}[htbp]
    \centering
    \begin{tikzpicture}[
        box/.style={draw, minimum width=1.8cm, minimum height=0.7cm, text centered, font=\small},
        arr/.style={->, >=latex, thick},
        input/.style={box, fill=gray!20},
        conv/.style={box, fill=blue!20},
        pool/.style={box, fill=green!20},
        fc/.style={box, fill=orange!20},
        output/.style={box, fill=yellow!20}
    ]
    
    \node[input] (input) {Input Image};
    \node[conv, below=0.3cm of input] (conv1) {Conv + ReLU};
    \node[pool, below=0.3cm of conv1] (pool1) {Max Pooling};
    
    \node[conv, right=0.8cm of pool1] (conv2) {Conv + ReLU};
    \node[pool, below=0.3cm of conv2] (pool2) {Max Pooling};
    \node[box, fill=gray!10, below=0.3cm of pool2] (flatten) {Flatten};
    
    \node[fc, right=0.8cm of pool2] (fc1) {Fully Connected};
    \node[output, below=0.3cm of fc1] (output) {Output (Softmax)};
    
    \draw[arr] (input) -- (conv1);
    \draw[arr] (conv1) -- (pool1);
    \draw[arr] (pool1) -- (conv2);
    \draw[arr] (conv2) -- (pool2);
    \draw[arr] (pool2) -- (flatten);
    \draw[arr] (flatten) -- (fc1);
    \draw[arr] (fc1) -- (output);
    
    \end{tikzpicture}
    \caption{Basic CNN architecture showing the typical sequence of convolutional, pooling, and fully connected layers used in image classification.}
    \label{fig:cnn_architecture}
\end{figure}

The hierarchical structure of CNNs is illustrated in Figure~\ref{fig:cnn_architecture}, which shows the typical sequence of layers in a convolutional neural network for image classification.

\subsection{CNNs for Image Classification}

CNNs have achieved state-of-the-art results in a wide range of image classification tasks, surpassing traditional machine learning algorithms in many benchmarks \citep{krizhevsky2012imagenet}. The ability of CNNs to automatically learn hierarchical features from raw pixel data makes them particularly well-suited for this task. In image classification, the goal is to assign a label to an image based on its content. CNNs accomplish this by extracting features from the image using convolutional and pooling layers, and then using fully connected layers to classify the image based on the extracted features.

The training process involves feeding the CNN a large dataset of labeled images and adjusting the parameters of the model to minimize the classification error. This is typically done using a gradient descent optimization algorithm, such as stochastic gradient descent (SGD) or Adam \citep{kingma2014adam}. The backpropagation algorithm is used to calculate the gradients of the loss function with respect to the model parameters, allowing the optimization algorithm to update the parameters in the direction that minimizes the loss \citep{goodfellow2016deep}.

Data augmentation techniques, such as random cropping, flipping, and rotating, are often used to increase the size of the training dataset and improve the generalization ability of the CNN \citep{krizhevsky2012imagenet}. These techniques help to prevent overfitting, which is a common problem when training deep learning models on limited datasets.

\subsection{Transfer Learning}

Transfer learning is a technique where knowledge gained from one task is applied to a different but related problem \citep{pan2009survey}. In malware detection using CNNs, this typically involves utilizing models pre-trained on large image datasets like ImageNet as a starting point for malware classification.

The process involves two main steps: First, starting with a network (such as VGG16 or ResNet50) that has been pre-trained on a large dataset, which has already learned to recognize general visual features. Second, adapting or fine-tuning this model on malware images. The early layers of these networks learn generic features like edges and textures that remain useful across different image classification tasks, while later layers can be retrained to identify malware-specific patterns \citep{yosinski2014transferable}.

This approach has proven effective in image-based malware detection, where pre-trained models have been successfully adapted to classify malware families represented as images \citep{Vasan2020IMCFNIM}. By leveraging transfer learning, malware classification systems can achieve higher accuracy with less training data and reduced computational resources, making it particularly suitable for practical implementation.

\subsection{CNN Architecture and Data Efficiency}

The relationship between CNN architectural complexity and data efficiency represents an important consideration for practical applications, particularly in domains where labeled data may be limited \citep{Sun2017RevisitingUE}. Research has shown that architectural choices significantly impact how models perform when training data is constrained.

More complex architectures with greater parameter counts typically require larger datasets to avoid overfitting, as they have increased capacity to memorize rather than generalize \citep{DBLP:journals/corr/abs-1902-07208}. Conversely, simpler architectures may generalize better from small datasets but might lack the representational capacity to capture complex patterns when more data is available. This creates an important trade-off that must be considered when selecting architectures for real-world applications.

The introduction of parallel pathways in dual-path architectures potentially impacts this trade-off in complex ways. While adding pathways increases the model's parameter count, it may also introduce beneficial architectural biases that improve generalization. For instance, a pre-trained pathway brings strong priors about visual features that may reduce the amount of task-specific data needed \citep{Vasan2020IMCFNIM}. The optimal balance between architectural complexity and data efficiency remains an active area of research, particularly in security domains where collecting large, labeled datasets presents significant challenges.

\subsection{Related Techniques}

While CNNs are the dominant architecture for image classification, other techniques exist, such as traditional machine learning algorithms (e.g., Support Vector Machines, Random Forests) and other types of neural networks (e.g., Recurrent Neural Networks). Traditional machine learning algorithms require manual feature engineering, which can be time-consuming and may not capture the most relevant features \citep{duda2012pattern}. Recurrent Neural Networks are designed for processing sequential data and are not well-suited for image classification \citep{elman1990finding}.

Other types of neural networks, such as Multilayer Perceptrons (MLPs), can also be used for image classification, but they typically do not perform as well as CNNs due to their lack of spatial awareness \citep{goodfellow2016deep}. MLPs treat each pixel in the image as an independent feature, ignoring the spatial relationships between pixels that are crucial for recognizing objects and patterns. The convolutional layers in CNNs are specifically designed to exploit the spatial structure of images, making them particularly effective for image-based malware detection \citep{nataraj2011malware, kalash2018}.

\section{Dual-Path CNN Architectures}
\subsection{Overview}
Dual-path CNNs augment traditional CNNs by employing multiple parallel pathways to process input data \citep{almasri2020dcmn}. These architectures are designed to capture distinct aspects or representations of the input, enabling the network to learn more comprehensive and robust features. In malware image classification, dual-path CNNs can process varied image types (e.g., bytecode and opcode) or extract different feature types (e.g., global structural and local semantic) from a single image. This multifaceted approach allows for a more nuanced understanding of malware, improving detection rates and reducing false positives.

The core principle of dual-path CNNs is to harness the complementary strengths of different CNN branches. Each branch can be tailored to focus on specific features or patterns, allowing the model to achieve a more complete representation. The features extracted from each branch are then combined using a feature fusion technique (discussed in Section \ref{section:feature_fusion}), enabling the model to learn relationships between the diverse feature sets. Thus, dual-path CNNs offer a framework for integrating varied information sources to enhance malware detection.

\subsection{Benefits of Dual-Path CNNs}
\begin{itemize}
    \item \textit{Improved Feature Extraction:} Dual-path CNNs extract more comprehensive features by processing diverse representations of the input data \citep{almasri2020dcmn}. This is particularly valuable for malware analysis, where different aspects of the binary code can provide insights into a malware's behavior and functionality. By analyzing both global structural and local semantic features, dual-path CNNs can achieve a more complete understanding.
    \item \textit{Enhanced Robustness:} Dual-path CNNs exhibit greater resilience to variations in the input data by learning features invariant to specific transformations or noise \citep{almasri2020dcmn}. For instance, a branch designed to extract global structural features may be less susceptible to local code obfuscation. This redundancy in feature extraction enhances the model's resilience to evasion attempts.
    \item \textit{Increased Accuracy:} By integrating features from multiple pathways, dual-path CNNs can achieve higher classification accuracy compared to single-path CNNs \citep{ Vasan2020IMCFNIM}. This stems from the model's ability to leverage the complementary strengths of different feature sets for more informed decisions. The synergistic effect of combining diverse features leads to improved performance.
    \item \textit{Training Data Considerations:} Research in deep learning has established that architectural complexity often correlates with training data requirements \citep{Sun2017RevisitingUE}. Dual-path CNN architectures, with their increased parameter counts, potentially face different training data dynamics compared to simpler architectures. Studies indicate that complex neural network architectures may require larger datasets to prevent overfitting and achieve optimal generalization \citep{DBLP:journals/corr/abs-1902-07208}. However, when incorporating transfer learning, pre-trained pathways can leverage features learned from large datasets, potentially reducing task-specific data requirements \citep{Vasan2020IMCFNIM}. The architectural separation of feature extraction into parallel specialized pathways creates distinct inductive biases that may alter the learning dynamics across different training data availability scenarios. Some research has observed that in certain contexts, multi-stream architectures exhibit different scaling behavior with respect to training data compared to single-stream approaches \citep{feichtenhofer2016convolutional}.
    
\end{itemize}

\subsection{Examples of Dual-Path CNN Architectures}
 \cite{almasri2020dcmn} introduced the Dual Convolutional Malware Network (DCMN), which combines a pre-trained ResNet-50 with a custom CNN. The ResNet-50 branch extracts high-level features, while the custom CNN captures structural details. This architecture demonstrates the efficacy of dual-path CNNs for malware image classification. Other examples include architectures processing different image representations (e.g., bytecode and opcode) in parallel or extracting distinct feature types (e.g., texture and entropy) from the same image. These varied architectures showcase the flexibility of dual-path CNNs for malware analysis.

\subsection{Connection to Feature Fusion: The Glue That Binds the Pathways Together} \label{section:feature_fusion}
Dual-path CNNs inherently depend on feature fusion to integrate information from the different pathways. The choice of feature fusion technique is critical for the dual-path CNN's performance \citep{Akilan2017, Boulahia2021}. As discussed in Section \ref{section:feature_fusion_techniques}, different techniques offer varying trade-offs in complexity, flexibility, and performance. Therefore, feature fusion is an integral component enabling the model to effectively synthesize information from multiple sources.

\subsection{Dual-Path CNN Architecture Visualization}

\begin{figure}[H]
        \centering
        \begin{tikzpicture}[
            box/.style={draw, minimum width=1.8cm, minimum height=0.7cm, text centered, font=\small},
            arr/.style={->, >=latex, thick},
            input/.style={box, fill=gray!20},
            vgg/.style={box, fill=blue!20},
            custom/.style={box, fill=green!20},
            fusion/.style={box, fill=purple!20},
            shared/.style={box, fill=gray!10},
            output/.style={box, fill=yellow!20}
        ]
        
        \node[input] (input) {Input Image};
        
        \node[vgg, below left=0.4cm and 0.6cm of input] (vgg1) {VGG16 Features};
        \node[custom, below right=0.4cm and 0.6cm of input] (custom1) {Custom CNN};
        
        \node[fusion, below right=0.8cm and -1.2cm of vgg1] (fusion) {Feature Fusion};
        
        \node[shared, below=0.4cm of fusion] (shared1) {Shared Layer 1};
        \node[shared, below=0.4cm of shared1] (shared2) {Shared Layer 2};
        
        \node[output, below=0.4cm of shared2] (output) {Output Layer};
        
        \draw[arr] (input) -- (vgg1);
        \draw[arr] (input) -- (custom1);
        \draw[arr] (vgg1) -- (fusion);
        \draw[arr] (custom1) -- (fusion);
        \draw[arr] (fusion) -- (shared1);
        \draw[arr] (shared1) -- (shared2);
        \draw[arr] (shared2) -- (output);
        
        \end{tikzpicture}
        \caption{Dual-path CNN architecture with intermediate feature fusion, combining a pre-trained VGG16 path with a custom CNN path. }
        \label{fig:dual_path_cnn}
    \end{figure}
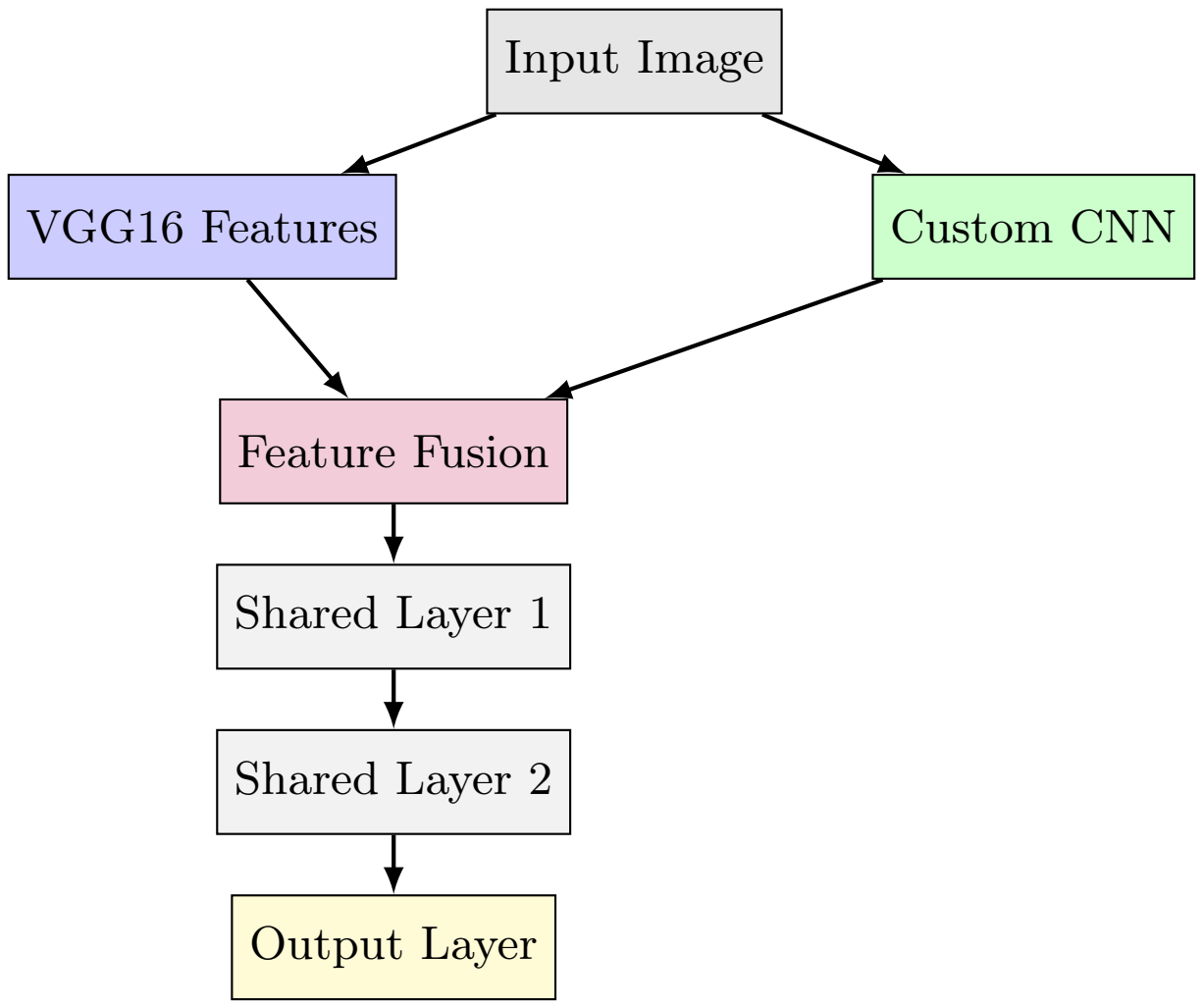

Figure~\ref{fig:dual_path_cnn} illustrates a typical dual-path CNN architecture with intermediate feature fusion, similar to the approach used in this thesis. The diagram shows how input images are processed through two parallel paths—each extracting complementary features—before being combined through feature fusion.
    
This architecture bridges the concepts discussed in previous sections, showing how dual paths extract different feature sets that are then combined through intermediate feature fusion before final classification.

\section{Feature Fusion Techniques} \label{section:feature_fusion_techniques}
\subsection{Overview}
Feature fusion is a crucial step in many machine learning applications, including image-based malware detection \citep{Boulahia2021}. It involves synthesizing features extracted from different sources or modalities to create a more comprehensive and informative data representation. In malware detection, feature fusion can combine features from different parts of the binary, distinct image representations, or even varied analysis techniques. Within dual-path CNNs, feature fusion is the mechanism by which information from different pathways is integrated for final classification.

The goal of feature fusion is to exploit the complementary strengths of different feature sets to improve the malware detection system's overall performance. By combining features capturing different aspects of the malware, the model can learn more robust and discriminative representations, leading to higher accuracy and fewer false positives. Thus, feature fusion is essential for maximizing the benefits of dual-path CNNs.

\subsection{Types of Feature Fusion}
Several approaches to feature fusion exist, each with its own advantages and disadvantages \citep{Boulahia2021}. Common techniques include:
\subsubsection{Early Fusion}

In early fusion, features are combined at the earliest stage, typically by concatenating them into a single feature vector. This approach is simple to implement but can be limited, as it doesn't allow the model to learn complex relationships between feature sets \citep{Boulahia2021}. Early fusion is often used when the raw modalities are combined ahead of feature extraction. However, it may not be optimal for dual-path CNNs, as it prevents fully exploiting the complementary strengths of different pathways.

\subsubsection{Intermediate Fusion} Intermediate fusion integrates features at a middle stage in the neural network, after each path has performed initial processing but before final classification. This approach enables the model to learn complex relationships between different feature sets while maintaining their distinct early-stage processing \citep{feichtenhofer2016convolutional}. Unlike early fusion where raw inputs are immediately combined, intermediate fusion first allows each pathway to extract specialized features independently.

The key advantage of this technique lies in its ability to facilitate non-linear interactions between complementary feature sets extracted from different network paths \citep{ramachandram2017deep}. For dual-path CNNs analyzing malware images, one path might excel at capturing global structural patterns while another identifies local textural details. When their features merge at an intermediate point, the network can leverage both perspectives simultaneously.

Feature map concatenation serves as the primary implementation method for intermediate fusion. This technique joins the feature maps along their channel dimension, preserving the complete information from both processing streams \citep{feichtenhofer2016convolutional}. After concatenation, subsequent network layers learn to identify which combinations of features from both paths provide the strongest signals for classification. Research suggests that the positioning of fusion within the network architecture is a critical design decision, with experimental results showing that fusion at mid-level convolutional blocks often yields superior performance \citep{feichtenhofer2016convolutional}. This balanced timing allows the network to benefit from both specialized processing in early layers and integrated analysis in later ones.

\subsubsection{Late Fusion}
Late fusion trains separate models on each feature set and combines their predictions \citep{Akilan2017, Boulahia2021}. This is flexible, allowing different models for each set, but can be complex and may not always achieve optimal performance. It's often used when the modality-wise classification results are combined. However, it may not be ideal for dual-path CNNs, as it prevents learning relationships between feature sets during training.

\subsection{Feature Fusion in Malware Detection}
In malware detection, feature fusion can combine features from different parts of the binary (header, code, data) or distinct image representations (grayscale, opcode, control flow graphs). By integrating these diverse sources, feature fusion can significantly improve detection systems.

Al-Masri et al. \citep{almasri2020dcmn} used feature fusion in DCMN by concatenating feature maps from ResNet-50 and custom CNN branches, leveraging both high-level features and structural details for improved performance. This demonstrates intermediate fusion's effectiveness in combining CNN branch strengths.

\subsection{Related Techniques}

While Dual-Path CNNs offer several advantages, other related techniques exist for improving the performance of CNNs. One such technique is the use of attention mechanisms, which allow the network to focus on the most relevant parts of the input image \citep{vaswani2017attention}. Attention mechanisms can be particularly useful for tasks where the input data contains irrelevant or noisy information. Another technique is the use of ensemble methods, which combine the predictions of multiple CNNs to improve accuracy and robustness \citep{lakshminarayanan2017simple}. Ensemble methods can be effective for reducing overfitting and improving generalization performance.

\chapter{Methodology}
\label{chapter:Methodology}

\section{Research Strategy}

The aim of this work is to investigate how varying training data sizes affect the comparative performance of single-path and dual-path CNN architectures in image-based malware classification. By comparing these architectures across different data availability conditions, this research seeks to determine which approach maintains better performance when training data is limited.

To address the research question, Experiment was chosen as the research strategy. 
Experiments are appropriate for this research as they allow for the systematic manipulation of independent variables in a controlled environment while observing their effects on dependent variables \citep{johannesson2014introduction}. This strategy enables direct comparison between different approaches and quantitative measurement of their performance differences across varying data constraints.

In this experiment, the study will evaluate how different CNN architectures perform under varying training data constraints for malware classification. The independent variables are (1) the network architecture (single-path VGG16 versus dual-path CNN with intermediate feature fusion) and (2) the training data size (20\%, 40\%, 60\%, and 80\% of available training data). For each training percentage, three separate training runs will be conducted using different random data subsets to ensure statistical robustness. The dependent variables include standard classification performance metrics (accuracy, precision, recall, F1-score), computational efficiency metrics (training time, model size), and robustness indicators (performance under noise perturbation).

Given the well-documented high accuracy of CNNs on image-based malware classification tasks. It is anticipated that potential performance saturation might occur where both architectures achieve near-perfect primary classification metrics.
Therefore, secondary metrics will also be analyzed including training efficiency, architectural complexity, and robustness to noise perturbation. These secondary metrics become critical differentiators when primary performance metrics reach saturation, providing deeper insights into architectural trade-offs that may not be apparent from classification accuracy alone.

To execute this research strategy, a dual-path CNN architecture will be implemented that combines a pre-trained VGG16 model with a custom CNN design, integrating their features through intermediate fusion. A baseline single-path VGG16 model will also be implemented for comparison. Both architectures will be trained on four different proportions of the training data. For each proportion p, multiple independent runs will be performed using StratifiedShuffleSplit, where p\% of the data forms the training set and the remaining (100-p)\% forms the corresponding test set for that specific run. This experimental design enables comprehensive comparison of how each architecture's performance scales with training data availability, with particular attention to the low-data regime (20-40\%) that represents realistic constraints in emerging threat detection.

For data collection, the study utilizes secondary data from the established Malimg dataset containing grayscale image representations of malware samples, which will undergo preprocessing to ensure compatibility with the architectures being tested.

For data analysis, quantitative analysis will be performed using standard classification metrics, complemented by efficiency metrics, complexity analysis, and robustness testing across multiple training runs and data sizes.

\section{Software Use and Implementation}

This study utilizes several established software libraries and frameworks to implement and evaluate the proposed dual-path CNN architecture. "For this implementation, Google Colab was selected, a cloud-based Jupyter notebook environment that provides free access to GPU resources. This platform was chosen for its accessibility, pre-installed machine learning libraries, and GPU acceleration capabilities, which are beneficial for training CNN models efficiently without requiring complex local setup.

The following key software libraries will be used in the implementation:

\begin{itemize} \item \textbf{TensorFlow and Keras:} TensorFlow is an open-source machine learning framework developed by Google, while Keras is a high-level neural networks API running on top of TensorFlow. These frameworks were selected because they provide comprehensive tools for building and evaluating deep learning models, offer efficient tensor operations with GPU acceleration, include extensive support for convolutional neural networks, and feature a user-friendly API that simplifies implementing complex architectures like dual-path networks.

\item \textbf{Scikit-learn:} This machine learning library will be used for calculating performance metrics (accuracy, precision, recall, F1-score), implementing stratified dataset splitting techniques, and performing statistical analysis across multiple training runs. Scikit-learn was chosen for its standardized implementation of evaluation metrics and robust data manipulation capabilities, ensuring reliable and comparable results across experiments.

\item \textbf{OpenCV:} This open-source computer vision library will be used for image resizing to standardize dimensions, pixel value normalization, and converting grayscale malware images to RGB format for VGG16 compatibility. OpenCV provides optimized implementations of these operations, making it suitable for preparing input data for CNN models.

\item \textbf{NumPy:} This fundamental package will be used for managing multi-dimensional arrays of image data, performing efficient numerical operations, handling data transformations during preprocessing, and calculating statistical measures across multiple training runs.

\item \textbf{Seaborn and Matplotlib:} These visualization libraries will be used for creating advanced visualizations of the results, particularly for generating performance versus data size curves with error bars, heatmap representations of confusion matrices, and statistical distribution plots. These libraries were selected for their specialized plotting functions that enable creation of informative and attractive statistical graphics.
\end{itemize}

The implementation will be written in Python, chosen for its extensive support for machine learning libraries, readability, and rapid development capabilities. Python is the standard language for deep learning research due to its comprehensive ecosystem of libraries and tools. This selection is justified by Python's widespread adoption in the machine learning community and the availability of well-maintained libraries that significantly reduce development time compared to other programming languages.

These software components interact in a structured workflow throughout the implementation. NumPy and OpenCV handle the initial data loading and preprocessing of malware images, preparing them for input to the neural networks. TensorFlow and Keras then serve as the core framework for building both the single-path and dual-path CNN architectures, managing the model training process, and implementing the feature fusion mechanism. During and after training, Scikit-learn processes the model outputs to calculate performance metrics, while matplotlib (using NumPy's numerical capabilities) generates visualizations of these metrics, training efficiency, and robustness behavior. Google Colab orchestrates this entire workflow, providing the computational environment and GPU acceleration for efficient model training and evaluation.

The implementation also incorporates efficient batch processing capabilities to handle the increased computational requirements of training multiple model instances across four different training data sizes with three runs per size. This expanded experimental design requires robust data management for creating and tracking multiple stratified dataset splits while maintaining reproducibility through consistent random seed management. The TensorFlow and scikit-learn libraries provide essential functionality for this purpose, allowing for systematic generation of multiple training subsets while preserving class distributions and ensuring experimental consistency.

\section{Data Collection}

In this work, secondary data collection was chosen as the method of data collection. Specifically, the study will use the Malimg dataset, which is a standard benchmark dataset in the domain of image-based malware analysis. This dataset consists of 9,339 malware samples converted to grayscale images, representing 25 different malware families. This publicly available dataset was chosen because it is widely recognized within the field of malware analysis research and has been extensively cited in the malware classification literature \citep{kalash2018}, making it an appropriate choice for this research as it allows for direct comparison with established approaches. The Malimg dataset was created by converting malware binaries into grayscale images using the technique proposed by \citep{nataraj2011malware}. Each byte of the malware binary is interpreted as a grayscale pixel value (0-255), and the bytes are arranged into a 2D array to form an image. The width of the image is fixed (typically at 256 pixels), while the height varies depending on the file size. This visualization technique preserves the structural patterns of the malware binary, allowing CNNs to identify visual signatures characteristic of specific malware families.

From the 25 families in Malimg, two visually distinct families were selected for a binary classification scenario: C2Lop.P (trojan) and Dialplatform.B (worm). The selection of these two specific malware families was based on four objective criteria to ensure impartiality. First, the class balance ratio (1:1.2) was quantitatively determined to minimize classification bias, with these two families having one of the closest ratios among all possible pairs in the dataset. Second, the taxonomic difference (trojan versus worm) represents distinct malware categories according to standard cybersecurity classification systems 
\citep{mitreattack2021}. Third, visual distinctiveness was assessed through basic image inspection and confirmed by prior research that has successfully classified these families using visual features \citep{kalash2018}. The different functionality of these malware types (data theft versus propagation) typically results in different code structures that manifest as visually distinct patterns when converted to images. Fourth, the sample count criterion ($>100$ samples per class) was established prior to family selection to ensure statistical validity. Additionally, these families have been objectively distinguished in previous research \citep{kalash2018}, providing external validation for their selection. Alternative family pairs were considered but rejected due to failing one or more of these objective criteria.

\begin{figure}[htbp] \centering \begin{subfigure}[b]{0.48\textwidth} \centering \includegraphics[width=\textwidth]{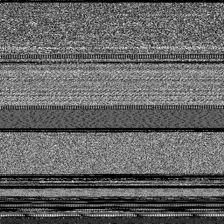} \caption{C2Lop.P (Trojan) malware visualized as grayscale image} \label{fig:c2lop} \end{subfigure} \hfill \begin{subfigure}[b]{0.48\textwidth} \centering \includegraphics[width=\textwidth]{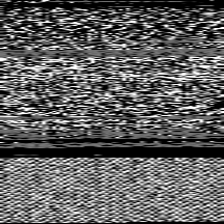} \caption{Dialplatform.B (Worm) malware visualized as grayscale image} \label{fig:dialplatform} \end{subfigure} \caption{Visual comparison of the two malware families selected for classification. Note the distinct structural patterns.} \label{fig:malware_comparison} \end{figure}

Moreover, the dataset's grayscale representation (as illustrated in Figure \ref{fig:malware_comparison}) is easily handled by standard CNN architectures—including pre-trained models such as VGG16—making it especially suitable for the transfer learning path in the dual-path CNN.

To prepare the dataset for model input, all images will be resized to 224×224 pixels so that they are compatible with the VGG16 architecture. Each image will then be normalized to the [0,1] range by dividing pixel values by 255, a common practice that helps stabilize training.

To implement the training data size variation, the entire dataset will be directly split into training and test sets using different proportions. Specifically, four experimental conditions will be created where training data comprises 20\%, 40\%, 60\%, and 80\% of the entire dataset, with corresponding test sets containing the remaining 80\%, 60\%, 40\%, and 20\% respectively. For statistical robustness, three distinct train-test splits will be generated for each percentage condition, using stratified random sampling with different random seeds. This stratified approach ensures that the class distribution remains consistent across all splits.

This design offers two methodological advantages: first, it provides a systematic way to evaluate how architecture performance varies with increasing training data availability; second, it inversely varies the test set size, providing more extensive evaluation data for models trained on smaller datasets (which are more likely to need robust testing). For example, models trained on just 20\% of the data will be evaluated on the largest test set (80\% of data), enabling more confident assessment of their generalization capabilities despite limited training exposure.

The selected dataset subset contains 323 samples in total (146 from C2Lop.P and 177 from Dialplatform.B), which is appropriate for this research for several reasons. First, the near 1:1.2 class ratio provides balanced representation that minimizes bias in the classification task. Second, when using transfer learning with pre-trained models like VGG16, smaller datasets can still yield reliable results as the model already contains generalized feature extractors developed from millions of images \citep{Vasan2020IMCFNIM}. Third, each class contains well over 100 samples, exceeding the minimum threshold often recommended for statistical validity in classification experiments \citep{vanderPloeg2014}. With this dataset size, the 20\% training condition will utilize approximately 65 samples for training (split proportionally between the two malware families), while the 80\% condition will use about 258 samples. Even at the smallest training size, each malware family maintains enough samples for meaningful model training \citep{DBLP:journals/corr/abs-1902-07208}, while the multiple training runs at each percentage enhance statistical reliability. The controlled experimental design further enhances the validity of results despite the moderate dataset size.

Alternative data collection methods were considered but not selected for this research. Creating a custom dataset by collecting and converting new malware samples would provide more recent examples but would require specialized security infrastructure, raise additional ethical concerns, and limit reproducibility. Generating synthetic malware images through techniques like generative adversarial networks could increase the dataset size but might introduce artificial patterns not representative of real malware. Other public datasets such as Microsoft Malware Classification Challenge dataset were considered, but many use feature vectors rather than raw images, making them less suitable for CNN-based visual analysis. The Malimg dataset was ultimately selected because it specifically provides grayscale image representations ideal for studying visual patterns, offers sufficient sample diversity, and enables comparability with existing research in image-based malware classification.

\section{Data analysis}

In this work, quantitative analysis of secondary data was chosen as the method of data analysis. Specifically, the research will evaluate how single-path and dual-path CNN architectures perform across varying training data sizes for malware classification. This approach allows for the assessment of not just which architecture performs better overall but which maintains better performance when training data is limited. A critical consideration for real-world deployment.

The following quantitative measures will be extracted from the experimental results:

\textbf{Primary Performance Metrics:}

1. \textit{Accuracy:} The proportion of correctly classified malware samples across both families, providing a straightforward measure of overall model performance. This widely-used metric will serve as the primary basis for comparing architecture performance.

2. \textit{F1-Score:} The harmonic mean of precision and recall, providing a balanced measure that accounts for both false positives and false negatives. This is particularly important for evaluating classification performance when class distributions are not perfectly balanced.

3. \textit{Precision and Recall:} Precision measures the proportion of true positive predictions among all positive predictions, while recall measures the proportion of true positives captured among all actual positives. These metrics provide insights into the balance between false positives and false negatives.

Recognizing the potential for performance saturation in standard classification metrics, particularly with highly distinguishable classes, several secondary metrics will also be analyzed that can differentiate architectures even when primary metrics reach near-perfect levels:

\textbf{Secondary Performance Indicators:}

1. \textit{Training Efficiency:} Measurement of training time across different data sizes and architectures. This metric is practically relevant for model development and retraining scenarios. The wall-clock time required to train each model configuration to convergence will be recorded and compared.

2. \textit{Model Complexity Analysis:} Quantification of architectural complexity through parameter counts (both trainable and non-trainable) and model file sizes. These metrics directly impact memory requirements, inference speed, and deployment feasibility on resource-constrained systems.

3. \textit{Robustness to Noise:} Assessment of model resilience to input perturbations by testing performance under increasing levels of Gaussian noise added to test images. This analysis simulates real-world scenarios where input data may be corrupted or slightly modified, providing insights into each architecture's generalization capabilities beyond the clean test set.

For data analysis, Python will be used with libraries such as scikit-learn for computing the classification metrics, and matplotlib and seaborn for creating visualizations. The AUC-ROC values will be calculated using scikit-learn's \verb|roc_auc_score| function, while confusion matrices will be generated using the \verb|confusion_matrix| function and visualized as heatmaps.

To minimize bias in the analysis, consistent preprocessing, training parameters, and evaluation methods will be used for both models across all training data sizes. The models will be trained on different subsets of the training data but evaluated on the corresponding test set generated for each specific training split to assess generalization performance across different data partitions. Additionally, confidence intervals and standard deviations will be reported to account for variance in model performance across different training runs.

\section{Alternative Research Strategies and Methods}

While Experiment was chosen as the research strategy, two alternative strategies could have been considered for investigating feature fusion in CNNs for malware classification:

\textbf{Design Science Research:} This strategy focuses on creating and evaluating artifacts to solve identified organizational problems \citep{johannesson2014introduction}. Under this approach, a dual-path CNN would have been developed as a novel artifact and evaluated its utility in addressing specific challenges in malware classification. However, this strategy was not selected because the primary focus is on understanding the impact of intermediate feature fusion within a specific architectural context rather than creating a solution for a particular organizational problem. This research aims to gain knowledge about the relationship between architectural design choices and model performance, which aligns more closely with the Experimental strategy.

\textbf{Comparative Analysis:} A comparative analysis approach would involve detailed comparison of multiple existing architectures without introducing experimental manipulations \citep{wohlin2012experimentation}. While this strategy could provide insights into the relative strengths of different approaches, it would not allow for the systematic manipulation of architectural elements (specifically, the addition of feature fusion) that is central to the research question. The experimental approach enables more direct causal inferences about the impact of feature fusion on model performance.

\subsubsection{Regarding alternative research methods, the following were considered:} 

\textbf{Feature Engineering Instead of Deep Learning:} Rather than using CNNs to automatically learn features from malware images, hand-crafted features could have been extracted from the images and used traditional machine learning classifiers such as Support Vector Machines or Random Forests. This method was not chosen because deep learning approaches have consistently outperformed traditional feature engineering methods for image classification tasks \citep{kalash2018}. Additionally, the research question specifically focuses on neural network architectures and intermediate feature fusion, which requires a deep learning approach. Furthermore, CNNs can automatically discover relevant features in the images, potentially identifying patterns that might be missed by manual feature engineering.

\textbf{Synthetic Data Generation:} Instead of using an existing dataset, we could have generated synthetic malware samples or augmented existing samples to create a larger and more diverse dataset. This approach was not adopted because creating realistic synthetic malware requires specialized expertise in malware development. Moreover, using standard, publicly available datasets allows for more direct comparison with existing research. Finally, the focus of the research is on architectural comparison rather than data augmentation techniques.

The chosen methods (experimental strategy with secondary data collection and quantitative analysis) provide the most direct approach to addressing the research question about the impact of intermediate feature fusion in dual-path CNNs for malware classification.

\section{Validity and Reliability}

\textbf{Validity:}

To ensure the validity of this experimental research, several measures were implemented addressing different aspects of validity:

For internal validity, identical preprocessing steps, training hyperparameters (such as learning rate, batch size, and optimization algorithm), and evaluation methods will be maintained for both architectures across all training data sizes. This controlled experimental setup ensures that any performance differences can be attributed to the architectural differences or training data size rather than confounding variables. For the training data size variation, stratified sampling will be used to maintain consistent class distributions across all training subsets. For each run, models trained on a specific percentage of data were evaluated on the corresponding complementary test set generated by the stratified split. This ensures that evaluation reflects generalization to unseen data from the same overall distribution for that specific partition, while the use of multiple runs and stratification addresses overall reliability. Additionally, the same random seed will be used for initializing model weights for corresponding runs across architectures, while using different random seeds for the three distinct training sets at each size percentage to ensure both consistency in comparisons and statistical robustness.

Regarding construct validity, standard classification metrics (accuracy, precision, recall, and F1-score) will be used that are widely accepted in the machine learning community as appropriate measures of classification performance. The secondary metrics (training efficiency, model complexity, and noise robustness) provide additional dimensions for comparison that are particularly valuable when primary metrics reach saturation. The Malimg dataset, being a standard benchmark in image-based malware analysis research, provides a valid representation of malware samples for the classification task. To further ensure construct validity, two malware families were selected with distinct visual characteristics but sufficient complexity to present a meaningful classification challenge.

To address external validity, we acknowledge that the binary classification between two specific malware families represents a simplified scenario compared to real-world malware classification challenges. However, this focused experiment serves as a proof of concept for comparing different architectural approaches. The selected training data sizes (particularly the 20\% and 40\% conditions) represent realistic scenarios that organizations might face when dealing with limited malware samples, especially for newer threats. Creating multiple distinct training sets for each data size percentage further enhances external validity by ensuring findings are not artifacts of a particular data subset. The inclusion of noise robustness testing adds another dimension to external validity by simulating real-world scenarios where input data may be imperfect or corrupted.

These validity measures collectively ensure that any observed performance differences can be reliably attributed to the architectural choices and training data size variations, which are the central focus of this investigation.

\textbf{Reliability:}

To ensure the reliability of this research, the following measures were implemented:

First, established, well-documented libraries (TensorFlow, Keras, scikit-learn) will be used for implementing the models and computing evaluation metrics. These libraries have undergone extensive testing and validation by the machine learning community, ensuring that the underlying calculations are standardized and reliable. The implementation will follow best practices for deep learning model training, including proper initialization, normalization, and optimization techniques.

Second, all data preprocessing steps will be clearly documented and implemented in a reproducible manner using version-controlled code. This includes consistent image resizing, normalization, and dataset splitting, ensuring that the same preprocessing pipeline is applied to all data. The specific preprocessing parameters (such as image dimensions and normalization factors) will be explicitly recorded.

Third, multiple training runs (three distinct runs) will be conducted for each data size percentage below 100 percent, calculating the average performance and standard deviation. This approach, aligned with standard scientific practice for ensuring statistical reliability, allows for accounting of variability introduced by specific data subset selection and random weight initialization. The use of multiple runs significantly increases confidence in the comparative analysis between architectures.

Finally, the experimental setup, including model architectures, hyperparameters, and evaluation procedures, will be thoroughly documented to enable reproduction of the experiments. All code and configuration files will be organized in a structured manner with appropriate documentation to ensure reproducibility.

These reliability measures ensure that the findings are consistent and reproducible, which is essential for building confidence in the conclusions drawn from this research.

\section{Ethical Considerations}

The ethical implications of malware classification research require careful consideration. While this research aims to contribute to cybersecurity knowledge, several ethical aspects have been addressed:

First, the Malimg dataset is used, which contains malware samples collected for research purposes. This dataset is publicly available and widely used in academic research, allowing for the study of malware patterns without creating or distributing harmful code. By using this established dataset, ethical concerns are avoided related to the creation or propagation of malware. Additionally, all research activities are conducted in an isolated, secure computing environment to prevent any risk of malware activation or distribution.

Regarding privacy considerations, this research does not involve personal data from individuals, as the Malimg dataset contains only malware binary samples converted to images. Nevertheless, it is recognized that malware can be used to compromise privacy, and thus this research indirectly contributes to privacy protection by advancing classification methods. Additionally, all results and findings are reported in aggregate form without revealing specific detection vulnerabilities that could be exploited to bypass security systems and compromise user privacy.

Second,While this research focuses on distinguishing between different types of malware, the knowledge gained could potentially inform both defensive and offensive cybersecurity practices. This dual-use potential is acknowledged and the focus is exclusively on the defensive applications in the discussion and conclusions. The primary intent of this research is to contribute to more effective malware classification techniques that can help protect information systems from cyber threats.

Furthermore, this study is transparent about the capabilities and limitations of the approach, avoiding exaggerated claims about its effectiveness or applicability. This honesty is important in cybersecurity research, where overstating the capabilities of classification techniques could lead to a false sense of security among users or organizations implementing similar approaches.

Lastly, all software and libraries used in this research are open-source or properly licensed, ensuring compliance with intellectual property rights. The research methodology adheres to the ethical guidelines for computer security research, focusing on responsible disclosure and the advancement of defensive capabilities rather than exploitative techniques.

\chapter{Main Results}
\label{chapter:MainResults}

\section{Technical Development}
For reproducibility purposes, all code used in this thesis is available at \url{https://github.com/Nooorab/malware-cnn-comparison}. A permanent link to the specific version used for these results can also be found in Software Heritage: \url{https://archive.softwareheritage.org/swh:1:rev:d33739cddcefcd15986cce8a93d8bd242756bd8f;origin=https://github.com/Nooorab/malware-cnn-comparison;visit=swh:1:snp:9f8f72eee5cd4fc51dbb758d040c2afad64c9ecc}.

This section provides a tutorial-like exposition of the implementation for comparing single-path and dual-path CNN architectures for malware image classification. This implementation was designed to be modular, reproducible and easily adaptable for future research.

\subsection{Experiment Configuration}

The parameters for the experimental setup were first defined to ensure a clear configuration of the architectures and the usage of the data set.
\begin{lstlisting}[language=Python, caption=Experiment configuration]
# Configuration Constants
SEED = 42
MALWARE_FAMILIES = ['C2LOP.P', 'Dialplatform.B'] # Target families
IMG_SIZE = (224, 224)  # Input dimension for VGG16 compatibility
TRAIN_PERCENTAGES= [0.2, 0.4, 0.6, 0.8] # Training percentages test
NUM_RUNS = 3   # Repetitions for each training percentage
Set seeds for reproducibility (Executed early in the script)
np.random.seed(SEED)
tf.random.set_seed(SEED)
random.seed(SEED)

\end{lstlisting}

\subsection{Data Loading and Preprocessing}

A data loading function was implemented that processes malware images and prepares them for use with CNN models. The pipeline handles loading grayscale images, resizing, conversion to RGB format (required by VGG16), and normalization:
\begin{lstlisting}[language=Python, caption={Core image loading and preprocessing steps.}] 
Inside loop iterating through image files...
img =cv2.imread(img_path, cv2.IMREAD_GRAYSCALE) # Load as grayscale
if img is not None:
img = cv2.resize(img, IMG_SIZE) # Resize to (224, 224)
# Convert to 3 channels for VGG16 compatibility
img = cv2.cvtColor(img, cv2.COLOR_GRAY2RGB)
img = img / 255.0 # Normalize pixel values to [0, 1]
images.append(img)
labels.append(family_idx) # Assign numerical label (0 or 1)

\end{lstlisting}

These processed images were stored in a NumPy array. Corresponding numerical labels (0 and 1) were assigned to each image based on its family, using a consistent mapping stored for reference.

\subsection{Model Architectures}

Two distinct CNN architectures were implemented and compared:

\subsubsection{Single-Path Model}
This model employed a transfer learning strategy using the VGG16 architecture pre-trained on ImageNet. The convolutional base of VGG16 was utilized as a fixed feature extractor by freezing all its layers (layer.trainable = False). A custom classification head was added on top of the VGG16 output, as shown below:

\begin{lstlisting}[language=Python, caption={Single-path CNN model implementation}] 
Assuming base_model is the frozen VGG16 base
base_model = VGG16(weights='imagenet', include_top=False,
input_shape=(224, 224, 3))
for layer in base_model.layers:layer.trainable=  False #Freeze base
x = base_model.output
x = Flatten()(x)
x = Dense(256, activation='relu')(x)
x = Dropout(0.5)(x) # Regularization
x = Dense(64, activation='relu')(x)
predictions = Dense(1, activation='sigmoid')(x) # Binary output

model = Model(inputs=base_model.input, outputs=predictions)


\end{lstlisting}

The head consists of flattening the features, followed by two Dense layers with ReLU activation and a Dropout layer for regularization. The final Dense layer uses a sigmoid activation function suitable for binary classification.

\subsubsection{Dual-Path Model}
This model implemented a more complex architecture combining features from both the pre-trained VGG16 network and a custom-built CNN path, using intermediate feature fusion.
\begin{itemize}
\item VGG16 Path: Similar to the single-path model, VGG16 pre-trained on ImageNet was used. However, fine-tuning was enabled for the later convolutional blocks (specifically block5) by setting only the earlier layers (vgg16.layers[:15]) as non-trainable. Features were extracted from the output of block4\_pool.

\item Custom CNN Path: A separate convolutional path was defined, consisting of multiple blocks of Conv2D, BatchNormalization, and MaxPooling2D layers, designed to process the input image independently and extract potentially complementary features.

\item Intermediate Fusion: The core of this architecture lies in the fusion step. Feature maps from the VGG16 path (block4\_pool output, shape approx. 14x14x512) and the custom CNN path (processed to a compatible shape, 14x14x512) were concatenated along the channel axis.
\end{itemize}
\begin{lstlisting}[language=Python, caption={Dual-path CNN model implementation}]
Assumes 'inputs' is the Keras Input Layer defined earlier
custom = inputs
filters = [32, 64, 128, 256, 512]
for f in filters:
  custom= Conv2D(f,(3,3),padding='same', activation='relu')(custom)
  custom = BatchNormalization()(custom)
  # Apply pooling except after the last Conv layer
  if f < filters[-1]:
    custom = MaxPooling2D((2, 2))(custom)

# --- Intermediate Feature Fusion ---
# Assumes 'vgg_out' holds features from VGG16's 'block4_pool'
fused = Concatenate()([vgg_out, custom])

# --- Post-Fusion Layers ---
x = Conv2D(512, (3, 3), activation='relu', padding='same')(fused)
x = BatchNormalization()(x)
x = MaxPooling2D((2, 2))(x)

# --- Classification Head ---
x = Flatten()(x)
x = Dense(256, activation='relu')(x)
x = Dropout(0.5)(x)
x = Dense(64, activation='relu')(x)
# The final Dense(activation='sigmoid') layer is added subsequently

\end{lstlisting}

This dual-path approach aims to leverage both general features from ImageNet (via VGG16) and potentially more domain-specific features learned by the custom path.

\subsection{Data Splitting Strategy}
\label{subsec:splitting_strategy_impl}
To ensure robust evaluation and address potential biases associated with a single fixed test set, a dynamic splitting strategy based on supervisor feedback was implemented using Scikit-learn's \texttt{StratifiedShuffleSplit}. 

For each experiment configuration defined by a training data percentage $p \in \{20\%, 40\%, 60\%, 80\%\}$, the entire available dataset (containing images from both malware families) was split $N=3$ times independently. 

In each split $i \in \{1,2,3\}$, $p\%$ of the data was randomly sampled for the training set $\left(X_{\text{train}}^{(p,i)},\ y_{\text{train}}^{(p,i)}\right)$, and the remaining $(100-p)\%$ formed the corresponding test set $\left(X_{\text{test}}^{(p,i)},\ y_{\text{test}}^{(p,i)}\right)$.

\begin{lstlisting}[language=Python, caption={Data Splitting Strategy}] 
Assumes 'base_images', 'base_labels', 'SEED', 'percentage', 'run' 
are defined
#Requires:from sklearn.model_selection import StratifiedShuffleSplit

# Calculate unique seed for reproducible splitting for this run
split_random_state = SEED + run + int(percentage * 100)

# Instantiate the splitter for this iteration
sss = StratifiedShuffleSplit(n_splits=1,
                             train_size=percentage,
                             random_state=split_random_state)

# Generate stratified train/test indices based on labels
train_idx, test_idx = next(sss.split(base_images, base_labels))

# Create the actual data subsets for this specific run
X_train_run = base_images[train_idx]
y_train_run = base_labels[train_idx]
X_test_run = base_images[test_idx]
y_test_run = base_labels[test_idx]
\end{lstlisting}

Stratification based on the malware family labels ($base\_labels$) was ensured by the StratifiedShuffleSplit object during the split operation, maintaining the proportion of samples from each family in both the training and testing sets for every individual split. This process yields a different, appropriately sized test set corresponding to each training set percentage and run, promoting a more comprehensive evaluation of generalization across various data partitions. Each split was made reproducible by using a unique random seed derived from a master seed (SEED=42), the run index i, and the training percentage p.

\subsection{Training Process}
\label{subsec:training_process}

The training for each model configuration (architecture type, training percentage, run number) 
was executed within a standardized function called 
\texttt{train\_\allowbreak and\_\allowbreak evaluate\_\allowbreak model}\hspace{0pt}.
A fresh model instance was created for each run.

Models were compiled using the Adam optimizer with Keras default parameters and binary cross-entropy loss, monitoring accuracy as the primary metric during training.

Training proceeded using \texttt{model.fit} on the specific training data split 

$\left(X_{\text{train}}\allowbreak ,\ y_{\text{train}}\right)$ 
for that run, with the corresponding test split 
$\left(X_{\text{test}}\allowbreak ,\ y_{\text{test}}\right)$ 
serving as validation data.

A batch size of 32 and a maximum limit of 50 epochs were set. 
The training process was regulated by three Keras callbacks:
\begin{lstlisting}[language=Python, caption={Keras callbacks used during model training.}]
callbacks = [
    # Saves the model with the best validation accuracy
    ModelCheckpoint(
        monitor='val_accuracy',    # Metric to monitor
        save_best_only=True,       # Save when 'monitor' improves
        save_weights_only=False,   # Save entire model structure
        mode='max',             # Improvement maximizing the metric
        verbose=1                  # Log messages when saving
    ),
    # Stops training early if validation accuracy stops improving
    EarlyStopping(
        monitor='val_accuracy',      # Metric to monitor
        patience=10,               # Epochs to wait before stopping
        restore_best_weights=True,   # Load best weights at the end
        verbose=1                    # Log when stopping occurs
    ),
    # Reduces learning rate if validation loss plateaus
    ReduceLROnPlateau(
        monitor='val_loss',         # Metric to monitor
        factor=0.5,            # Factor by which LR is reduced 
        patience=5,       # wait for improvement before reducing LR
        min_lr=1e-6,           # Lower bound for the learning rate
        verbose=1              # Log when LR is reduced
    )
]


\end{lstlisting}

The ModelCheckpoint callback saved the complete model corresponding to the epoch with the highest validation accuracy directly to the designated Google Drive path. EarlyStopping monitored validation accuracy and terminated training if no improvement was observed for 10 consecutive epochs, restoring the model weights from the best epoch to prevent overfitting and reduce computation time. ReduceLROnPlateau adaptively decreased the learning rate by half if the validation loss stagnated for 5 epochs, potentially aiding convergence.

\subsection{Evaluation Metrics}
\label{subsec:evaluation_metrics_impl}

Following the methodology outlined, a comprehensive set of metrics was collected and calculated using Scikit-learn to evaluate and compare the models:

\begin{itemize}
\item Primary Classification Metrics: Accuracy, F1-Score (chosen for balance), AUC-ROC (Area Under the Receiver Operating Characteristic Curve, chosen for threshold independence), Precision, and Recall.

\item Efficiency: Total training time (\texttt{\detokenize{time.time}}), inference time per image (measured via \texttt{\detokenize{model.predict}} timing), parameter counts (\texttt{model.\-count\_params}), and saved model file size (\texttt{os.\-path.\-getsize}).

\item {Loss:} Minimum validation loss achieved during training. (\textit{Extracted from Keras history })
    
\item Robustness: Accuracy on test images corrupted by additive Gaussian noise (\texttt{numpy.\-random.\-normal}).

\end{itemize}

These metrics enabled a multi-faceted comparison, essential for differentiating the models when primary performance metrics were saturated. The subsequent section presents the aggregated results from these evaluations.

\section{Experimental Results}

This section presents the empirical findings from comparing the single-path and dual-path CNN architectures, following the experimental design detailed in Chapter 3. The research question sought to understand how varying training data size affects the comparative performance of these architectures. As anticipated, primary classification metrics reached saturation early, necessitating a broader evaluation. Therefore, "comparative performance" is assessed here across multiple dimensions: not only standard classification accuracy but also training efficiency, robustness to noise, and model complexity, providing a holistic view of architectural suitability under the tested conditions. The following subsections detail the results for each of these performance aspects.

\subsection{Primary Classification Performance}

Table \ref{tab:performance_metrics} summarizes the mean primary classification metrics across the three runs for each training data percentage. Both the single-path and dual-path architectures achieved perfect scores (1.0) for accuracy, precision, recall, and F1-score across all tested training data sizes, from 20\% to 80\%. The AUC-ROC metric, also calculated, similarly yielded a perfect score of 1.0 in all cases (results shown in analysis logs but omitted from the table for brevity as it provided no additional discriminatory information). This saturation indicates that the classification task between the 'C2LOP.P' and 'Dialplatform.B' families using visual features is highly separable and relatively straightforward for both CNN architectures, even when trained on significantly limited data (20\%). Consequently, these primary performance metrics alone do not differentiate between the two architectures in this specific experimental setup, highlighting the necessity of examining the secondary indicators.

\begin{table}[htbp]
\centering
\caption{Mean classification metrics across different training data percentages.}
\label{tab:performance_metrics}
\begin{tabular}{lccccc}
\hline
\textbf{Model Type} & \textbf{Training \%} & \textbf{Accuracy} & \textbf{Precision} & \textbf{Recall} & \textbf{F1-Score} \\
\hline
Single-Path & 20\% & 1.0 & 1.0 & 1.0 & 1.0 \\
 & 40\% & 1.0 & 1.0 & 1.0 & 1.0 \\
 & 60\% & 1.0 & 1.0 & 1.0 & 1.0 \\
 & 80\% & 1.0 & 1.0 & 1.0 & 1.0 \\
\hline
Dual-Path & 20\% & 1.0 & 1.0 & 1.0 & 1.0 \\
 & 40\% & 1.0 & 1.0 & 1.0 & 1.0 \\
 & 60\% & 1.0 & 1.0 & 1.0 & 1.0 \\
 & 80\% & 1.0 & 1.0 & 1.0 & 1.0 \\
\hline
\end{tabular}
\end{table}

\subsection{Training Efficiency}
\begin{figure}[htbp]
\centering
\includegraphics[width=\textwidth]{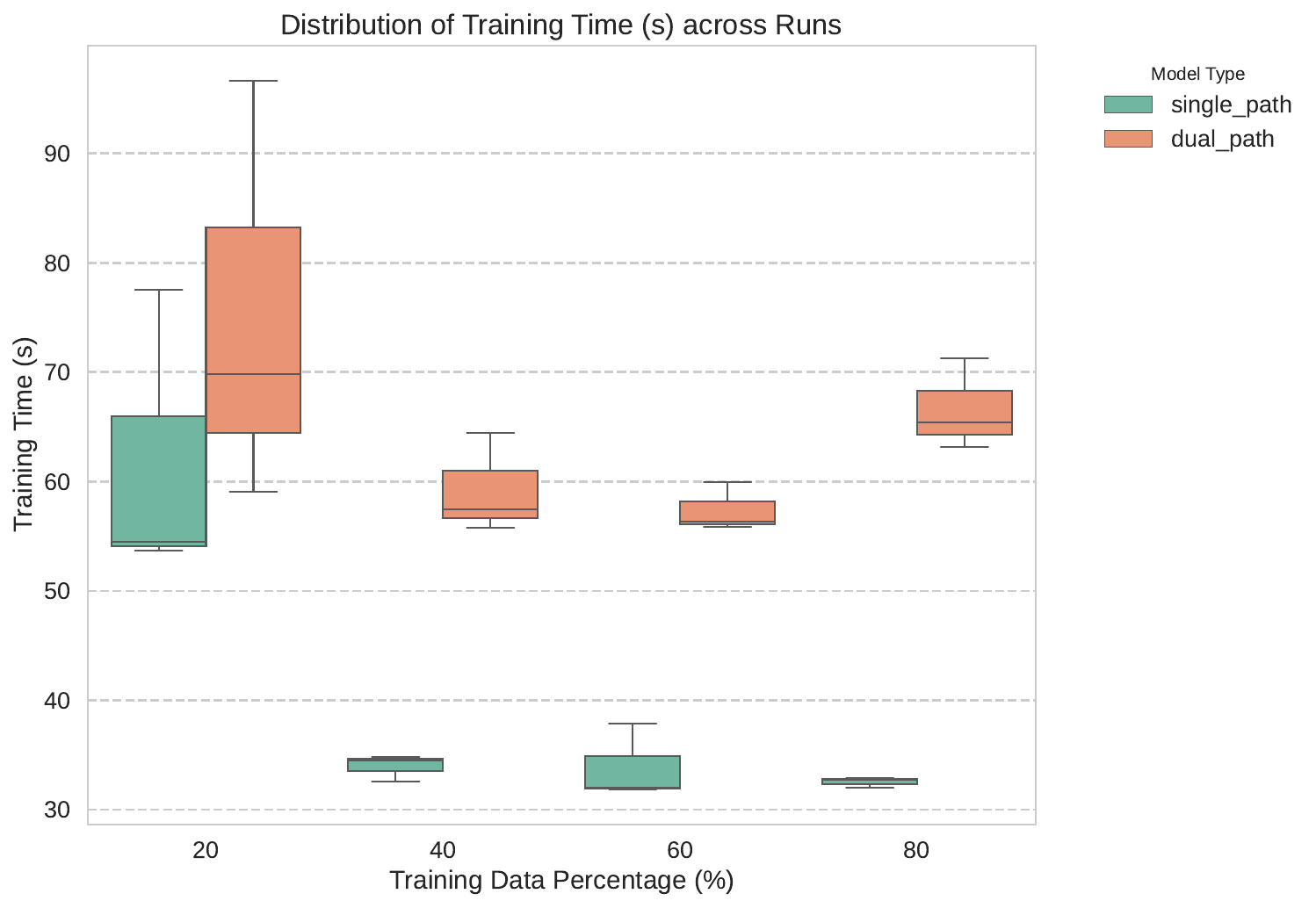}
\caption{Distribution of training times (seconds) across 3 runs for single-path vs. dual-path models at different training data percentages.}
\label{fig:training_time}
\end{figure}

Training efficiency, measured by the wall-clock time required to train each model until convergence (dictated by the EarlyStopping callback), reveals a clear distinction between the architectures. Figure \ref{fig:training_time} presents boxplots illustrating the distribution of training times for both architectures across the three runs at each training data percentage. Each box represents the interquartile range (IQR) of the training times for the three runs, with the line inside marking the median. The whiskers typically extend to show the range of the data (excluding potential outliers, though none appear prominent here). The plot visually confirms that the single-path model (represented by, e.g., blue boxes) consistently exhibits lower median training times and generally less variability (smaller box sizes) compared to the dual-path model (e.g., orange boxes) across all data percentages.

Table \ref{tab:training_efficiency} quantifies the average training times and standard deviations, confirming the trend observed in Figure \ref{fig:training_time}. The single-path model demonstrates consistently lower mean training times. At 20\% training data, the single-path model trains on average 13.29 seconds faster (a 17.7\% speed advantage relative to the dual-path time), and this advantage increases significantly with more data. At 80\% training data, the single-path model is 34.08 seconds faster, representing a 51.2\% speed advantage. This highlights substantial computational savings achieved by the simpler architecture, particularly relevant as dataset sizes grow, without any compromise in classification accuracy for this task.

\begin{table}[htbp]
\centering
\caption{Mean training efficiency comparison (Mean Time ± Std Dev across 3 runs). Speed advantage calculated as (Time\_Dual - Time\_Single) / Time\_Dual.}
\label{tab:training_efficiency}
\begin{tabular}{lccc}
\hline
\textbf{Training Data \%} & \textbf{Single-Path (s)} & \textbf{Dual-Path (s)} & \textbf{Speed Advantage} \\
\hline
20\% & 61.89 ± 13.53 & 75.18 ± 19.35 & 17.7\% faster \\
40\% & 33.94 ± 1.22 & 59.25 ± 4.59 & 42.7\% faster \\
60\% & 33.91 ± 3.44 & 57.41 ± 2.23 & 40.9\% faster \\
80\% & 32.52 ± 0.48 & 66.60 ± 4.18 & 51.2\% faster \\
\hline
\end{tabular}
\end{table}

\subsection{Robustness to Noise}

\begin{figure}[htbp]
\centering
\includegraphics[width=\textwidth]{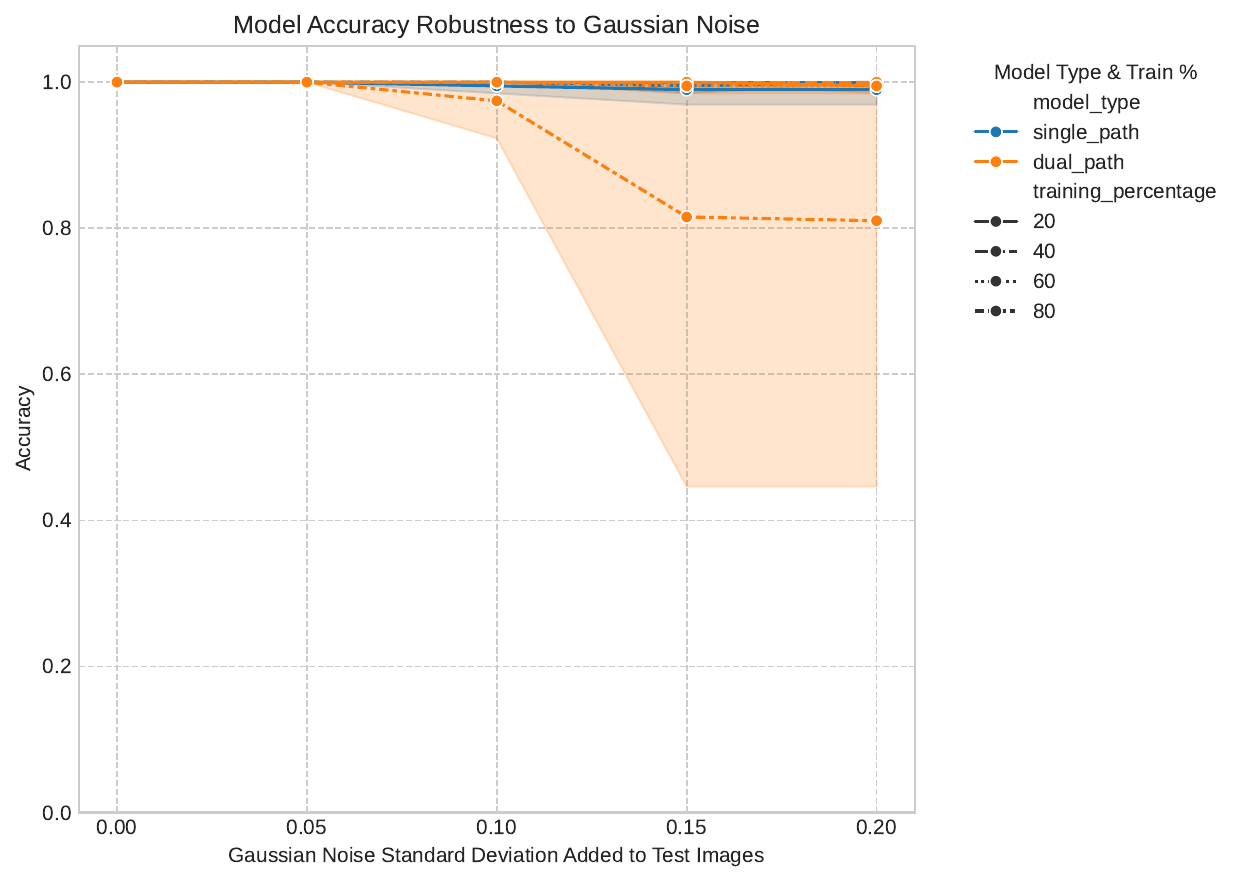}
\caption{Model accuracy under increasing Gaussian noise levels. Lines show mean performance across all runs; shaded areas indicate variability.}
\label{fig:robustness}
\end{figure}

Model robustness was evaluated by introducing additive Gaussian noise with increasing standard deviation (0.05 to 0.20) to the test images and measuring the impact on classification accuracy. Figure \ref{fig:robustness} presents the performance of the two model architectures under these conditions.

The Figure \ref{fig:robustness} displays two main trend lines: the blue solid line represents the average accuracy of the single-path model, and the orange dashed line represents the average accuracy of the dual-path model. These averages are calculated across all training percentages (20-80\%) and all three experimental runs for each noise level. The shaded areas around each line depict the variability (such as standard deviation or a 95\% confidence interval) associated with that average.

Observing the trends, both model types achieve perfect average accuracy on clean data (0.0 noise) and show minimal degradation at a low noise level (0.05). However, their robustness diverges as noise intensity increases. The single-path model's average accuracy (blue line) remains consistently high, dipping only marginally below 1.0 even at the highest noise standard deviation of 0.20. Furthermore, the narrow blue shaded band indicates low variability in its performance across the different conditions.

In contrast, the dual-path model's average accuracy (orange dashed line) declines more sharply, particularly between noise levels 0.10 and 0.15, reaching an average of approximately 0.81 at the 0.20 noise level. The significantly wider orange shaded band at higher noise levels highlights substantial variability in the dual-path model's accuracy under noise, suggesting inconsistent performance across different runs or training data sizes.

This visual comparison is quantified by the summary data in Table \ref{tab:robustness}, which details the mean and standard deviation at the maximum noise level (0.20) for different training percentages. The table confirms the trend seen in the figure, particularly the large performance drop and high variance for the dual-path model trained on 80\% data (mean accuracy 0.810 ± 0.315) compared to the perfect accuracy of the single-path model (1.000 ± 0.000).

\begin{table}[htbp]
\centering
\caption{Mean Accuracy ± Std Dev under maximum noise level (0.20 std dev).}
\label{tab:robustness}
\begin{tabular}{lcc}
\hline
\textbf{Training Data \%} & \textbf{Single-Path} & \textbf{Dual-Path} \\
\hline
20\% & 0.990 ± 0.018 & 0.995 ± 0.009 \\
40\% & 1.000 ± 0.000 & 1.000 ± 0.000 \\
60\% & 1.000 ± 0.000 & 0.995 ± 0.009 \\
80\% & 1.000 ± 0.000 & 0.810 ± 0.315 \\
\hline
\end{tabular}
\end{table}

Collectively, Figure \ref{fig:robustness} and Table \ref{tab:robustness} demonstrate that the single-path architecture possesses superior average robustness to additive Gaussian noise compared to the dual-path architecture within this experimental context.

\subsection{Model Complexity Analysis}

Finally, the internal complexity of the models was examined by counting their parameters (the internal 'settings' they learn) and checking the size of the saved model files. Table \ref{tab:complexity} summarizes this.

\begin{table}[htbp]
\centering
\caption{Model architecture complexity comparison.}
\label{tab:complexity}
\begin{tabular}{lcccc}
\hline
\textbf{Architecture} & \textbf{Total Parameters} & \textbf{Trainable} & \textbf{Non-trainable} & \textbf{File Size} \\
\hline
Single-Path & 21,153,985 & 6,439,297 & 14,714,688 & 129.92 MB \\
Dual-Path & 20,368,257 & 12,729,985 & 7,638,272 & 174.99 MB \\
\hline
\end{tabular}
\end{table}

Table \ref{tab:complexity} shows two key things. First, trainable vs. non-trainable: The single-path model uses many 'frozen' (non-trainable) settings from VGG16 and only learns about 6.4 million new settings. The dual-path model freezes fewer VGG16 settings but learns many more (about 12.7 million) because it adjusts parts of VGG16 and learns the settings for its custom path.

Second the file Size: The dual-path model, having learned more unique settings, resulted in significantly larger saved files (175 MB vs 130 MB).

\begin{figure}[H]
\centering
\includegraphics[width=0.8\textwidth]{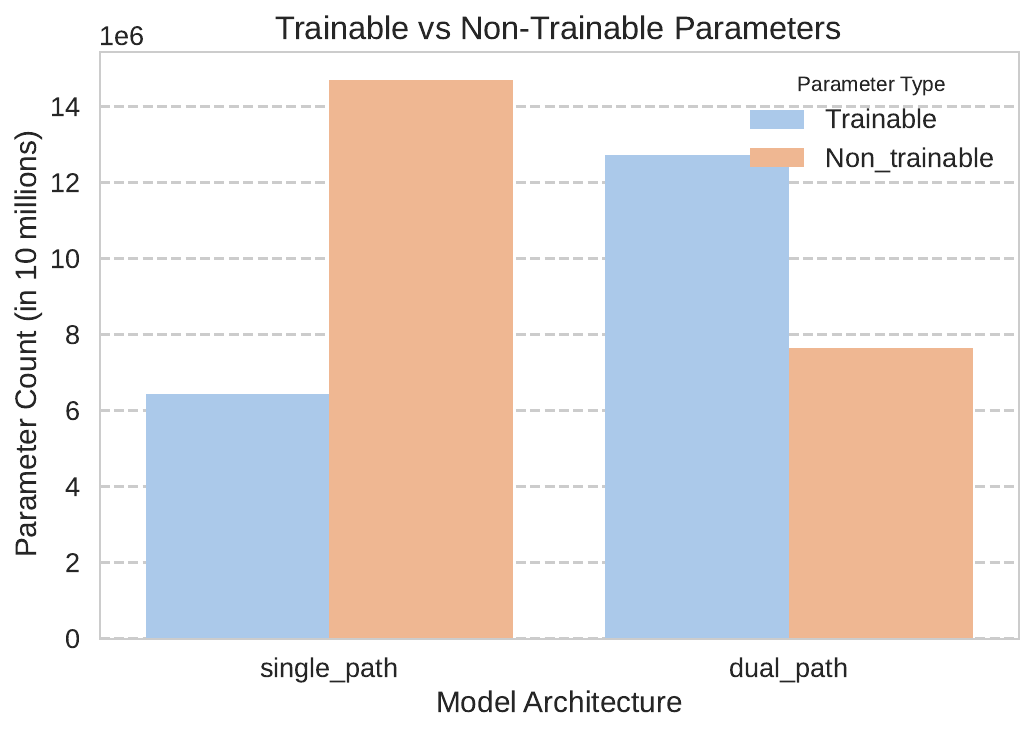}
\caption{Comparison of Trainable vs. Non-trainable parameters.}
\label{fig:parameter_breakdown}
\end{figure}

Figure \ref{fig:parameter_breakdown} visually confirms the difference in parameter types using bars, where blue represents trainable settings and orange represents non-trainable settings.

The larger number of trainable parameters in the dual-path model is the primary reason it took longer to train (Table \ref{tab:training_efficiency}) and resulted in larger files (Table \ref{tab:complexity}).

\subsection{Summary of Findings}

The experimental results yield several key insights:

1. Both architectures achieve perfect classification performance across all metrics and training data sizes.

2. The single-path architecture provides substantial computational advantages:
18-51\% faster training time and (approx. 35\% smaller file size) due to fewer trainable parameters.

3. Most surprisingly, the single-path architecture exhibits superior robustness to noise at higher training percentages, maintaining perfect accuracy where the dual-path model degrades significantly.

These findings challenge the assumption that more complex architectures necessarily yield better results in visual malware classification. The simpler single-path CNN architecture represents a more efficient and robust solution for distinguishing between C2LOP.P and Dialplatform.B malware families.

\chapter{Conclusion}
\label{chapter:Conclusion}

This work investigates the problem of image-based malware classification using Convolutional Neural Networks (CNNs), focusing on the practical constraint of limited training data availability. More specifically, the research considers the question of how varying training data size affects the comparative performance of single-path and dual-path CNN architectures with intermediate feature fusion in malware classification using grayscale image representations. The primary goal was to determine which architectural approach offers better resilience and overall suitability when training data is scarce.

To answer this research question, two distinct CNN architectures were implemented. A single-path model using transfer learning with VGG16 and a dual-path model combining VGG16 with a custom CNN path. The study conducted 24 experiments (2 architectures × 4 training percentages × 3 runs) using malware images from two families (C2LOP.P and Dialplatform.B), and analyzed the results across multiple performance dimensions.

The experiments yielded clear, albeit somewhat unexpected, results. Both the single-path and dual-path architectures achieved perfect classification performance (1.0 for Accuracy, F1-Score, AUC-ROC) across all tested training data sizes, indicating high separability for this specific binary task. Consequently, the comparison shifted to secondary metrics. Results showed that the single-path architecture was significantly more efficient, requiring 18-51\% less training time and producing model files approximately 26\% smaller than the dual-path architecture. Furthermore, the single-path model demonstrated markedly superior robustness when Gaussian noise was added to the test images, maintaining near-perfect accuracy under conditions where the dual-path model's performance degraded substantially, particularly when trained on more data.

The most remarkable conclusion from this study is that, for this highly separable binary classification task, the architecturally simpler single-path model provided superior overall performance when considering efficiency and robustness, despite the identical perfect accuracy scores. This challenges the assumption that increased complexity always yields better results in this domain.

\section{Related Work}

This research contributes to the field of image-based malware classification, building upon foundational work by \cite{nataraj2011malware} (visualization), \cite{kalash2018} (CNN effectiveness on Malimg), and \cite{Vasan2020IMCFNIM} (transfer learning/fine-tuning). While confirming the general effectiveness of CNNs, this study provides specific insights into architectural trade-offs under varying data availability, a dimension less explored previously.

The findings complement related work such as \cite{almasri2020dcmn}, who also explored dual-path networks (ResNet-50 + custom CNN with concatenation) but focused on accuracy maximization on the full Malimg dataset rather than the systematic analysis of performance trade-offs (efficiency, robustness) across varying training data sizes conducted here. This work also differs from multi-modal approaches like MalNet \cite{yan2018detecting}, which combined CNNs on images with LSTMs on opcode sequences using late fusion. This study maintained focus on the image modality to specifically isolate the effects of architectural choices (single-path vs. dual-path intermediate fusion) under varying data availability, revealing significant differences in efficiency and robustness even when accuracy was saturated. These results add a practical dimension to the discussion initiated by \cite{DBLP:journals/corr/abs-1902-07208} regarding complexity versus performance in data-constrained scenarios within the specific context of malware visualization.

\section{Delimitations}

The conclusions drawn are subject to certain delimitations impacting their generalizability:

\begin{itemize}
\item Dataset Scope: The experiments were conducted solely on two specific malware families (C2LOP.P, Dialplatform.B) selected from the Malimg dataset. While chosen based on objective criteria (Section 3.3), these two families proved highly separable visually, leading to performance saturation. The results may not generalize to multi-class scenarios involving all 25 Malimg families or other families with less distinct visual features.
\item Dataset Age: The Malimg dataset, while standard, represents malware samples from an earlier timeframe (around 2011). Modern malware might employ different structures or obfuscation techniques not fully represented in this dataset.
\item Specific Architectures: Only one specific implementation of a single-path (frozen VGG16 head) and dual-path (fine-tuned VGG16 block5 + specific custom CNN + intermediate concatenation) architecture was tested. Different base models (e.g., ResNet, EfficientNet), custom path designs, or fusion techniques (e.g., attention, addition) might yield different comparative results.
\item Robustness Testing Scope: Robustness was only evaluated against additive Gaussian noise. Performance against other perturbations (e.g., adversarial attacks, different noise types, image transformations) was not assessed.
\end{itemize}
These delimitations mean the conclusion – that the simpler single-path model was superior here – is strongly tied to the specific context of this experiment (two easily separable families, Gaussian noise). The relative merits of the architectures might change significantly on more complex, larger-scale tasks.

\section{Ethical implications}

The findings of this research have several ethical and societal implications related to malware detection and cybersecurity.

Improved malware classification techniques, especially those that are computationally efficient, can enhance cybersecurity defenses and protect users from harmful software. By demonstrating that simpler architectures can achieve comparable or superior performance to more complex ones, this research contributes to making effective malware detection more accessible and deployable, potentially reducing the impact of malicious software on society.

However, it is important to acknowledge that techniques developed for malware detection could potentially be studied and circumvented by malware developers. This creates an adversarial dynamic where malware might evolve to avoid detection by visual analysis methods. The finding that noise can reduce classification accuracy of certain models (as shown in the robustness analysis) could potentially be exploited to evade detection.

Furthermore, the binary nature of malware classification involves ethical considerations around false positives and false negatives. While this study achieved perfect classification, real-world applications would need to balance the harm caused by misclassifying legitimate software as malware against the risk of failing to detect actual malware.

Finally, the computational efficiency gained through simpler architectures has positive environmental implications by reducing the energy consumption and carbon footprint associated with training and deploying malware detection systems.

\section{Considerations for future work.}

This research points towards several promising directions for future investigation:
 
Evaluate the single-path vs. dual-path comparison on the full Malimg dataset or newer, larger multi-class malware image datasets, incorporating methods to handle class imbalance.

Explore variations in architecture: different pre-trained bases (e.g., ResNet), alternative custom CNN designs, and different intermediate fusion techniques (e.g., attention mechanisms).

Conduct more extensive robustness analysis using diverse noise types, image corruptions, and potentially adversarial examples tailored for malware images.

Investigate why the single-path model exhibited greater robustness through feature analysis and visualization techniques.

\section{Usage of AI Tools}

Grammarly tool was used during the writing process for correcting spelling and grammar. Perplexity was then used to find relevant literature. GitHub Copilot was used with code implementation and debugging.

\bibliographystyle{agsm}
\bibliography{bibliographyFiles/bibliography}
\addtocontents{toc}{\bigskip}
\addcontentsline{toc}{part}{Bibliography}

\appendix
\cleardoublepage
\addtocontents{toc}{\bigskip}
\addcontentsline{toc}{part}{Appendices}

\end{document}